\documentclass[a4paper,10pt]{article}
\pdfoutput=1
\usepackage{jheppub}
\usepackage{slashed}
\usepackage[dvipsnames,table]{xcolor}
\usepackage{hyperref} 
\usepackage{xspace}
\usepackage[tight]{subfigure}
\usepackage{amsmath}
\usepackage{amssymb}
\usepackage{amsfonts}
\usepackage{mathrsfs}
\usepackage{comment}
\usepackage{afterpage}
\usepackage{verbatim}
\usepackage{booktabs}
\usepackage{array}
\usepackage[title,titletoc]{appendix}
\usepackage{tabularx}
\newcolumntype{C}[1]{>{\centering\arraybackslash}p{#1}}

\def\reffi#1{\mbox{Fig.~\ref{#1}}}
\def\reffis#1{\mbox{Fig.~\ref{#1}}}

\def\refse#1{\mbox{Sect.~\ref{#1}}}

\def\citere#1{\mbox{Ref.~\cite{#1}}}
\def\citeres#1{\mbox{Refs.~\cite{#1}}}

\newcommand{\newc}{\newcommand}
\newc{\beq}{\begin{equation}}
\newc{\eeq}{\end{equation}}
\newc{\beqn}{\begin{eqnarray}}
\newc{\eeqn}{\end{eqnarray}}
\newc{\bit}{\begin{itemize}}
\newc{\eit}{\end{itemize}}
\newc{\ben}{\begin{enumerate}}
\newc{\een}{\end{enumerate}}
\newc{\bce}{\begin{center}}
\newc{\ece}{\end{center}}
\newc{\bfi}{\begin{figure}}
\newc{\efi}{\end{figure}}

\newcommand{\ri}{\mathrm i}

\newcommand{\rd}{\mathrm d}

\newcommand{\rT}{{\mathrm{T}}}

\newcommand{\rL}{{\mathrm{L}}}

\newcommand{\ie}{\emph{i.e.}\ }
\newcommand{\eg}{\emph{e.g.}\ }

\newcommand{\GeV}{\ensuremath{\,\text{GeV}}\xspace}

\newcommand{\fb}{{\ensuremath\unskip\,\text{fb}}\xspace}

\newcommand{\PH}{\ensuremath{\text{H}}\xspace}
\newcommand{\Pj}{\ensuremath{\text{j}}\xspace}
\newcommand{\Pp}{\ensuremath{\text{p}}}
\newcommand{\Pe}{\ensuremath{\text{e}}\xspace}

\newcommand{\Pt}{\ensuremath{\text{t}}\xspace}

\newcommand{\Pg}{\ensuremath{\text{g}}}

\newcommand{\PW}{\ensuremath{\text{W}}\xspace}
\newcommand{\PZ}{\ensuremath{\text{Z}}\xspace}

\newcommand{\Mt}{\ensuremath{m_\Pt}\xspace}
\newcommand{\MH}{\ensuremath{M_\PH}\xspace}
\newcommand{\MWOS}{\ensuremath{M_\PW^\text{OS}}\xspace}
\newcommand{\MW}{\ensuremath{M_\PW}\xspace}
\newcommand{\MZOS}{\ensuremath{M_\PZ^\text{OS}}\xspace}
\newcommand{\MZ}{\ensuremath{M_\PZ}\xspace}

\newcommand{\GZ}{\ensuremath{\Gamma_\PZ}\xspace}
\newcommand{\GZOS}{\ensuremath{\Gamma_\PZ^\text{OS}}\xspace}

\newcommand{\GWOS}{\ensuremath{\Gamma_\PW^\text{OS}}\xspace}

\newcommand{\alphas}{\ensuremath{\alpha_\text{s}}\xspace}

\newcommand{\MVOS}{\ensuremath{M_{V}^\text{OS}}\xspace}%
\newcommand{\GVOS}{\ensuremath{\Gamma_{V}^\text{OS}}\xspace}%

\newcommand{\order}[1]{\ensuremath{\mathcal{O}{\left(#1\right)}}\xspace}

\newcommand{\Sherpa}{{\sc Sherpa}\xspace}

\newcommand{\OpenLoops}{{\sc OpenLoops}\xspace}

\newcommand{\madgraph}{{\sc\small MadGraph}\xspace}

\newcolumntype{.}{D{.}{.}{-1}}
\newcolumntype{d}[1]{D{.}{.}{#1}}
\colorlet{tableoverheadcolor}{gray!37.5}
\colorlet{tableheadcolor}{gray!25}
\colorlet{tablerowcolor}{gray!12.5}

\def\draftdate{\relax}
\def\mda{\relax}
\def\mua{\relax}
\def\mla{\relax}
\def\draft{
\def\thtystars{******************************}
\def\sixtystars{\thtystars\thtystars}
\typeout{}
\typeout{\sixtystars**}
\typeout{* Draft mode!
         For final version remove \protect\draft\space in source file *}
\typeout{\sixtystars**}
\typeout{}
\def\draftdate{\today}
\def\mua{\marginpar[\boldmath\hfil$\uparrow$]%
                   {\boldmath$\uparrow$\hfil}\color{black}%
                    \typeout{marginpar: $\uparrow$}\ignorespaces}
\def\mda{\color{red}\marginpar[\boldmath\hfil$\downarrow$]%
                   {\boldmath$\downarrow$\hfil}%
                    \typeout{marginpar: $\downarrow$}\ignorespaces}
\def\mla{\marginpar[\boldmath\hfil$\rightarrow$]%
                   {\boldmath$\leftarrow $\hfil}%
                    \typeout{marginpar: $\leftrightarrow$}\ignorespaces}
\def\Mua{\marginpar[\boldmath\hfil$\Uparrow$]%
                   {\boldmath$\Uparrow$\hfil}\color{black}%
                    \typeout{marginpar: $\uparrow$}\ignorespaces}
\def\Mda{\color{red}\marginpar[\boldmath\hfil$\Downarrow$]%
                   {\boldmath$\Downarrow$\hfil}%
                    \typeout{marginpar: $\downarrow$}\ignorespaces}
\def\Mla{\marginpar[\boldmath\hfil\textcolor{red}{$\Rightarrow$}]%
                   {\boldmath\textcolor{red}{$\Leftarrow $}\hfil}%
                    \typeout{marginpar: $\leftrightarrow$}\ignorespaces}
\overfullrule 5pt
\oddsidemargin 15mm
\marginparwidth 29mm
}

\newcommand{\mc}{\mathcal}

\newcommand{\pt}[1]{p_{\rT,{#1}}}
\newcommand{\nnb}{\nonumber}

\newcommand{\rU}{{\rm U}}

\newcommand{\noun}[1]{{\scshape #1}}
\newcommand{\POWHEG}{\noun{Powheg}}

\newcommand{\pwhg}{\noun{Powheg-Box}}
\newcommand{\moca}{\noun{MoCaNLO}}
\newcommand{\bbmc}{\noun{BBMC}}
\newcommand{\sher}{\noun{Sherpa}}
\newcommand{\mulb}{\noun{MulBos}}
\newcommand{\madg}{\noun{MG5\_aMC}}
\newcommand{\stri}{\noun{STRIPPER}}
\newcommand{\red}{\color{blue}}

\title{Precise Standard-Model predictions for polarised Z-boson pair production and decay at the LHC}

\author[a]{Costanza Carrivale,}
\author[b]{Roberto Covarelli,}
\author[c]{Ansgar Denner,}
\author[d]{Dongshuo Du,}
\author[c]{Christoph Haitz,}
\author[e]{ Mareen Hoppe,}
\author[f]{Martina Javurkova,}
\author[g]{Duc Ninh Le,}
\author[h]{Jakob Linder,}
\author[f]{Rafael Coelho Lopes de Sa,}
\author[i]{Olivier Mattelaer,}
\author[j]{Susmita Mondal,}
\author[k]{Giacomo Ortona,}
\author[k,1]{Giovanni Pelliccioli,\note{Coordinators}}
\author[l,1]{Rene Poncelet,}
\author[m]{Karolos Potamianos,}
\author[l]{Richard Ruiz,}
\author[n]{Marek Sch\"onherr,}
\author[e]{Frank Siegert,}
\author[d]{Lailin Xu,}
\author[d]{Xingyu Wu,}
\author[h]{Giulia Zanderighi}

\affiliation[a]{University of Perugia, Department of Physics and Geology and INFN, 06123 Perugia, Italy}
 \affiliation[b]{University of Torino, Department of Physics and INFN, 10125 Torino, Italy} 
  \affiliation[c]{Universit\"at W\"urzburg, Institut f\"ur Theoretische Physik und Astrophysik, 97074 W\"urzburg, Germany} 
  \affiliation[d]{University of Science and Technology of China, Department of Modern Physics and State Key Laboratory of Particle Detection and Electronics, Hefei 230026, China}
  \affiliation[e]{Dresden University of Technology, Institute for Nuclear and Particle Physics, D–01062 Dresden, Germany} 
  \affiliation[f]{University of Massachusetts, Department of Physics, Amherst, MA 01003-4525, United States } 
  \affiliation[g]{Phenikaa University, Phenikaa Institute for Advanced Study, Hanoi 12116, Vietnam} 
  \affiliation[h]{Max-Planck-Institut f\"ur Physik and Technische Universit{\"a}t M{\"u}nchen, Physik-Department, 85748 Garching, Germany} 
  \affiliation[i]{Universit\'e Catholique de Louvain, Centre for Cosmology, Particle Physics and Phenomenology, B-1348 Louvain-la-Neuve, Belgium} 
 \affiliation[j]{University of Wisconsin, Department of Physics, Madison, WI 53706-1390, United States } 
 \affiliation[k]{University of Milano--Bicocca, Department of Physics and INFN, 20126 Milano, Italy}
  \affiliation[l]{Institute of Nuclear Physics, 31–342 Krakow, Poland} 
   \affiliation[m]{University of Warwick, Department of Physics, Coventry, CV4-7AL, United Kingdom}
  \affiliation[n]{Institute for Particle Physics Phenomenology, Department of Physics, Durham University, Durham DH1-3LE, United Kingdom}

\emailAdd{giovanni.pelliccioli@unimib.it, rene.poncelet@ifj.edu.pl}
\date{\draftdate}
\preprint{COMETA-2025-16, IFJPAN-IV-2025-10, IPPP/25/22, MCNET-25-07}
\abstract{
Providing accurate theoretical predictions in the Standard Model for processes with polarised electroweak bosons is crucial to understand more in-depth the electroweak-symmetry breaking mechanism and to enhance the sensitivity to potential new-physics effects. Motivated by the rapidly increasing number of polarisation analyses of di-boson processes with LHC data, we carry out a comprehensive study of the inclusive production of two polarised Z bosons in the decay channel with four charged leptons.
We perform a detailed comparison of fixed-order predictions obtained with various Monte Carlo programs which rely on different signal-definition strategies, assessing non-resonant and interference effects by contrasting polarised results with unpolarised and full off-shell ones. 
For the first time, we accomplish the combination of NNLO QCD and NLO EW corrections, setting the new state-of-the-art perturbative accuracy for polarised Z-boson pairs at the LHC.
The impact of parton-shower matching and multi-jet merging predictions is investigated by scrutinising calculations obtained with event generators that are typically used in experimental analyses. 
Integrated and differential results are discussed in a realistic fiducial setup and compared to publicly available ATLAS results. 
}

\begin{document}
\strut\hfill
\maketitle

\section{Introduction}\label{sec:intro}

The experimental data collected during Run-2 and Run-3 operational stages of the Large Hadron Collider (LHC) allows for precise measurements of processes involving two electroweak (EW) gauge bosons. The upcoming High-Luminosity phase will enable even higher precision in this direction. Isolating the polarisation modes of EW bosons is crucial to broaden our understanding of the mechanism of electroweak-symmetry breaking (EWSB). In the Standard Model (SM), the EW bosons acquire a mass and a longitudinal-polarisation state via the EWSB mechanism. Consequently, any deviation in the production rate of longitudinally polarised bosons would point out effects beyond the Standard Model (BSM). Therefore, analysing EW-boson polarisations at TeV-scale energies serves as a tool to distinguish between SM and BSM dynamics. For a sound interpretation of LHC data in terms of boson polarisations, theoretical predictions in the SM (and beyond) need to rely on a stable definition of polarised signals, to achieve the highest precision possible and to include realistic effects to permit direct comparison with experimental data.

Polarisation measurements have been carried out with Run-2 data by means of polarised-template fits in inclusive $\PZ\PW$ \cite{Aaboud:2019gxl,CMS:2021icx,ATLAS:2022oge,ATLAS:2024qbd} and $\PZ\PZ$ production \cite{ATLAS:2023zrv}, and same-sign $\PW\PW$ scattering \cite{Sirunyan:2020gvn,ATLAS:2025wuw}.

A number of phenomenological studies have explored processes involving polarised EW bosons at the LHC in various channels \cite{Ballestrero:2017bxn,BuarqueFranzosi:2019boy,Ballestrero:2019qoy,Ballestrero:2020qgv,Denner:2020bcz,Denner:2020eck,Poncelet:2021jmj,Denner:2021csi,Pellen:2021vpi,Le:2022lrp,Le:2022ppa,Denner:2022riz,Dao:2023pkl,Hoppe:2023uux,Pelliccioli:2023zpd,Denner:2023ehn,Dao:2023kwc,Javurkova:2024bwa,Dao:2024ffg,Denner:2024tlu,Grossi:2024jae}. Simulating precisely polarised bosons is especially challenging due to radiative corrections that impact both their production and decays. To date, the SM predictions beyond leading order (LO) have been computed for inclusive di-boson production \cite{Denner:2020bcz,Denner:2020eck,Poncelet:2021jmj,Denner:2021csi,Le:2022lrp,Le:2022ppa,Denner:2022riz,Dao:2023pkl,Hoppe:2023uux,Pelliccioli:2023zpd,Denner:2023ehn,Dao:2023kwc,Dao:2024ffg,Grossi:2024jae}, $\PW$+jet production \cite{Pellen:2021vpi}, and vector-boson scattering \cite{Denner:2024tlu}. 

In several multi-boson processes, longitudinal-polarisation contributions are characterised by significant cancellations amongst pure-gauge and Higgs-mediated diagrams. These delicate cancellations could be spoiled by BSM effects, potentially leading to sizeable cross-section enhancements. BSM effects in polarisation studies have been explored in a limited number of production mechanism, with a focus on vector-boson scattering (VBS) that is commonly regarded as the gold-plated channel for an in-depth investigation of the EWSB mechanism.
Studies of polarisation observables have been carried out also for inclusive di-boson production in the presence of anomalous couplings or effective operators \cite{Rahaman:2018ujg,Baglio:2019uty,Rahaman:2019lab,ElFaham:2024uop,Thomas:2024dwd}. 

Machine-learning techniques have also been explored to extract polarisation fractions as well as to achieve polarisation tagging \cite{Searcy:2015apa,Lee:2018xtt,Lee:2019nhm,Grossi:2020orx,Kim:2021gtv,Li:2021cbp,Grossi:2023fqq}.

In inclusive production of two polarised bosons, next-to-leading order (NLO) QCD and NLO EW corrections are known for all production mechanism in the fully leptonic decay channel \cite{Denner:2020bcz,Denner:2020eck,Denner:2021csi,Le:2022lrp,Le:2022ppa,Dao:2023pkl,Denner:2023ehn,Dao:2023kwc,Dao:2024ffg}, NLO QCD corrections have been computed for $\PZ\PW$ in the semileptonic channel as well \cite{Denner:2022riz}, while next-to-next-to-leading order (NNLO) QCD corrections are known only for $\PW^+\PW^-$ production with leptonic decays \cite{Poncelet:2021jmj}. The matching to parton shower (PS) has been achieved at approximate \cite{Hoppe:2023uux} and exact \cite{Pelliccioli:2023zpd} NLO QCD accuracy in the \sher~and \pwhg~frameworks, respectively. Approximate NLO merging is also available in the \sher~generator \cite{Hoppe:2023uux}. Including PS effects in publicly available Monte Carlo codes is presently restricted to the modelling of initial-state QCD radiation, while decays are either modelled at fixed NLO or at LO matched to leading-logarithmic QED showers. Predictions for loop-induced, gluon-initiated polarised di-boson production are only available at LO \cite{Denner:2020bcz,Poncelet:2021jmj,Denner:2021csi,Dao:2023kwc,Javurkova:2024bwa}. 

The combination of \sloppy NNLO QCD and NLO EW corrections, known for the off-shell processes \cite{Kallweit:2019zez} also matched to PS \cite{Lindert:2022qdd}, is still missing in the literature for the case of polarised intermediate bosons. Additionally, no comprehensive comparison amongst independent calculations exists yet, especially concerning matched predictions. Moreover, polarised predictions obtained by different groups lack homogeneity and general recommendations for the LHC community.
This work aims to achieve these purposes by comparing and combining several calculations, both at fixed order and including realistic effects from PS matching and multi-jet merging. 

This work is focused on the production of polarised $\PZ$-boson pairs in the fully leptonic decay channel with two pairs of different flavour charged leptons. In \refse{subsec:setup} we present the calculation details. The SM input parameters and the fiducial setup considered for comparisons and phenomenological analyses are provided in \refse{subsec:setupspec}, while the technical details of used Monte Carlo generators are given in \refse{subsec:mctools}. The numerical results are presented in \refse{sec:results}. The validation of polarised cross sections at fixed order and a phenomenological analysis in the presence of parton-shower matching and merging are depicted respectively in \refse{sec:fo_results} and \refse{sec:ps_results}, including both integrated and differential cross sections. A tailored discussion on polarisation fractions is showcased in \refse{sec:fractions}. A comparison between state-of-the-art SM predictions and ATLAS-simulation results used for a recent Run-2 analysis \cite{ATLAS:2023zrv} is presented in \refse{sec:comp_exp}. We draw our conclusions in \refse{sec:concl}.

\section{Details of the calculations}\label{subsec:setup}
\begin{figure}[ht]
  \centering
  \subfigure[Doubly resonant\label{fig:ZZres}]{\includegraphics[width=0.25\textwidth]{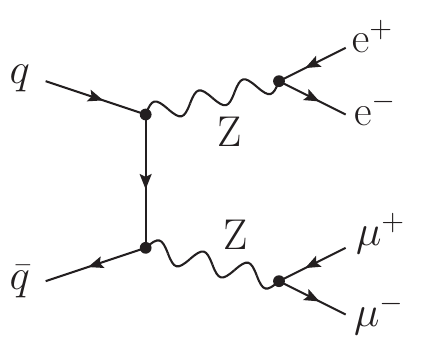}}
  \subfigure[Photon mediated\label{fig:nonZres}]{\includegraphics[width=0.25\textwidth]{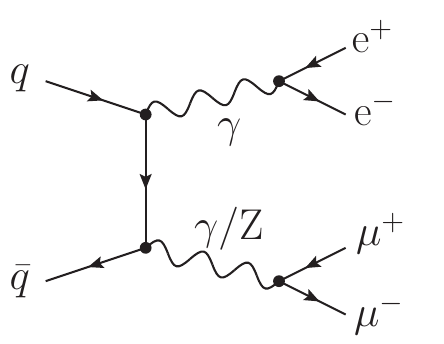}}
  \subfigure[Nested \label{fig:periph}]{\includegraphics[width=0.3\textwidth]{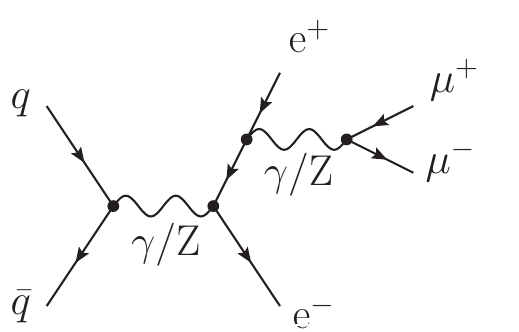}}
  \caption{Representative tree-level diagrams contributing to the off-shell process $q\bar{q}\rightarrow\Pe^+\Pe^-\mu^+\mu^-$. Only resonant topologies \ref{fig:ZZres} are retained in the DPA and NWA, while photon-mediated \ref{fig:nonZres} and nested \ref{fig:periph} contributions are dropped.}  \label{fig:LOdiags}
\end{figure}

In this work, the off-shell production of four charged leptons at the LHC is considered,
\begin{eqnarray}\label{eq:fullprocess}
\Pp\Pp & \,\, \longrightarrow \,\,\Pe^+\Pe^-\,\mu^+\mu^-+X\,.
\end{eqnarray}
This process receives both resonant and non-resonant contributions at any order in perturbation theory. 
Sample diagrams are shown in \reffi{fig:LOdiags}.
In order to define a signal for bosons with definite physical polarisation modes, the target bosons have to be on mass-shell. This requires matrix elements in a factorised form (production $\times$ decays). To achieve this, non-resonant contributions like those depicted in \reffis{fig:nonZres}--\ref{fig:periph} need to be dropped, while retaining only (doubly) resonant diagrams (\reffi{fig:ZZres}), with intermediate on-shell $\PZ$ bosons in $s$-channel. In formulas, the (un)polarised signals are defined for the process,
\begin{eqnarray}\label{eq:resprocess}
\Pp\Pp & \,\, \longrightarrow \,\,\PZ_{\lambda} \,\, (\rightarrow\Pe^+\Pe^-)&\,\,\PZ_{\lambda'}\,\,(\rightarrow\mu^+\mu^-)+X\,,
\end{eqnarray}
where the polarisation states $\lambda, \lambda'$ can be longitudinal ($\rL$) or transverse ($\rT$), and 
if the label $\rU$ is used, the corresponding $\PZ$ boson is meant to be unpolarised\footnote{While in all polarised predictions of this work the same polarisation state of an intermediate boson is selected for the amplitude and its corresponding complex conjugate, \ie diagonal terms of the spin-density matrix are evaluated, it is possible to give a theoretically sound definition of off-diagonal terms as well.}.
At the level of squared matrix elements, the transverse polarisation state ($\rT$) will be considered in this work, defined as the coherent sum of the left- and right-handed polarisation terms (including left-right interference).

The on-shellness of the intermediate bosons can be achieved in two different manners: the pole approximation 
\cite{Stuart:1991cc,Stuart:1991xk,Stuart:1995ba,Aeppli:1993rs,Denner:2000bj,Denner:2005fg,Denner:2019vbn}, 
dubbed double-pole approximation (DPA) in the case of two resonances, 
and the narrow-width approximation (NWA) 
\cite{Richardson:2001df,Uhlemann:2008pm,Artoisenet:2012st}.
Originally exploited to simplify off-shell calculations and especially the evaluation of loop amplitudes, 
in the context of polarisation analyses at the LHC they enable to define polarised signals \cite{Ballestrero:2017bxn,BuarqueFranzosi:2019boy,Ballestrero:2019qoy,Ballestrero:2020qgv,Denner:2020bcz,Denner:2020eck,Poncelet:2021jmj,Denner:2021csi,Le:2022lrp,Le:2022ppa,Denner:2022riz,Dao:2023pkl,Hoppe:2023uux,Pelliccioli:2023zpd,Denner:2023ehn,Dao:2023kwc,Dao:2024ffg,Denner:2024tlu,Grossi:2024jae},
which are currently used in
template fits of experimental data \cite{Aaboud:2019gxl,Sirunyan:2020gvn,CMS:2021icx,ATLAS:2022oge,ATLAS:2023zrv,ATLAS:2024qbd}. 
The two approaches feature the same intrinsic accuracy compared to off-shell calculations, \emph{i.e.} $\mc O (\Gamma/M)$, 
but their technical details are different.
\begin{itemize}
    \item[$\bullet$] In the NWA \cite{Richardson:2001df,Uhlemann:2008pm,Artoisenet:2012st}, 
    the momenta are generated for the production process in the first place, then the on-shell bosons are decayed to obtain the kinematics for the complete process. The production-process amplitude is evaluated with the on-shell, production-level kinematics, and then multiplied by the corresponding amplitudes for the decays.
   The matrix elements are obtained by squaring the factorised amplitudes, retaining where needed spin correlations by means of the calculation of the full spin-density matrix. Partial off-shell effects are recovered by means of a smearing of final-state momenta according to Breit-Wigner distributions for the intermediate bosons, leading to off-shell phase-space weights. 
    \item[$\bullet$] In the DPA \cite{Aeppli:1993rs,Denner:2000bj,Denner:2005fg,Denner:2019vbn}, momenta are generated for the complete process with general off-shell kinematics for the intermediate bosons. The amplitude for the off-shell process is expanded in the complex plane about the pole mass of the bosons (appearing in the propagator denominators) and only the leading terms are retained. The numerator of the resonant amplitude is evaluated with on-shell-projected momenta for the external particles, obtained by modifying the original kinematics in such a way that the intermediate bosons are on-shell. Squared amplitudes feature automatically complete spin correlations. The propagator denominators are evaluated with the original off-shell kinematics. The phase-space weight is also evaluated with the original off-shell kinematics.
    
\end{itemize}
 \noindent   
In both approaches, assigning a physical polarisation state to the two intermediate $\PZ$ bosons ($\lambda$ and $\lambda'$),
the cross section is proportional to,
\begin{eqnarray}\label{eq:master}
     && \left|{\mc A}^{(\lambda,\lambda')}\right|^2 \,\rd{\rm \Phi}_4(\hat{s};{k_{1\ldots4}}) \,=\, \\
      &=&\left|\frac{\left({\mc P}_{\mu\nu}(\bar{k}_{12},\bar{k}_{34})\, \varepsilon_\lambda^{\mu}(\bar{k}_{12})\,\varepsilon_{\lambda'}^{\nu}(\bar{k}_{34})\right)\,\left(\varepsilon_\lambda^{*\alpha}(\bar{k}_{12}){\mc D}_{\alpha}(\bar{k}_{1},\bar{k}_{2})\right)\,\left(\varepsilon_{\lambda'}^{*\beta}(\bar{k}_{34}){\mc D}_{\beta}(\bar{k}_{3},\bar{k}_{4})\right)}{\left(k^2_{12}-\MZ^2+\ri\GZ\right)\left(k^2_{34}-\MZ^2+\ri\GZ\right)}\right|^2\rd{\rm \Phi}_4(\hat{s};{k_{1\ldots4}})\;, \nnb 
    \end{eqnarray}
where $\mc P$ is the truncated production-level amplitude and $\mc D$ the decay one. The (un)barred kinematics understands (off)on-shell momenta of intermediate $\PZ$ bosons. $k_{12} = k_1+k_2$ and $k_{34} = k_3+k_4$ are the off-shell momenta of the intermediate $\PZ$ bosons, while $\bar{k}_{12}$ and $\bar{k}_{34}$ are the corresponding on-shell momenta. The differences between the DPA and the NWA lie in how the off-shell and on-shell kinematics are generated. In both cases, gauge invariance is guaranteed by the evaluation of the numerator in Eq.~\ref{eq:master} with on-shell kinematics of the bosons and leptons.

The NWA and DPA do not represent the only strategies to achieve this purpose. 
For example, the default decay-chain approach in  \madg~(dubbed BW) relies on resonant-diagram selection, without any on-shell approximation. Since this could easily lead to gauge dependence of the results, the numerical sampling of the resonance off-shell-ness can be restricted to be close to the pole mass of the resonance, up to a certain number of widths. This strategy, up to possible subtleties related to gauge dependence to be checked case by case, is very simple and widely used in experimental analyses \cite{Aaboud:2019gxl,Sirunyan:2020gvn,CMS:2021icx,ATLAS:2022oge,ATLAS:2023zrv,ATLAS:2024qbd}.

All proposed strategies to define polarised templates in the NWA, DPA and BW approaches \cite{Ballestrero:2017bxn,BuarqueFranzosi:2019boy,Ballestrero:2019qoy,Ballestrero:2020qgv,Denner:2020bcz,Denner:2020eck,Poncelet:2021jmj,Denner:2021csi,Pellen:2021vpi,Le:2022lrp,Le:2022ppa,Denner:2022riz,Dao:2023pkl,Hoppe:2023uux,Pelliccioli:2023zpd,Denner:2023ehn,Dao:2023kwc,Dao:2024ffg,Denner:2024tlu,Grossi:2024jae}
 rely on the replacement
\beq\label{eq:amp_replacement}
\sum_{\lambda'}\varepsilon^\mu_{\lambda'}\varepsilon^{\nu*}_{\lambda'} \,\longrightarrow\,\varepsilon^\mu_{\lambda}\varepsilon^{\nu*}_{\lambda}\,, \qquad \lambda = \rL\,(\textrm{longitudinal}), \pm\,(\textrm{right/left-handed})\,,
\eeq
to single out in the propagator numerator an individual physical polarisation state $\lambda$ of an intermediate $\PZ$ boson (with momentum $p$) 
from the sum over all physical and un-physical polarisation terms. 
At variance with the replacement introduced in Eq.~\ref{eq:amp_replacement} (often dubbed propagator-truncation method), a novel strategy has been recently proposed to extract polarised signals~\cite{Javurkova:2024bwa} by directly modifying Feynman rules used in the amplitude construction.
The weak-boson fields are redefined at the level of Feynman rules as sums of states with definite polarisation state which act
as separate propagators. Polarised cross sections, including partial off-shell effects, are obtained via diagram selection and squaring the amplitude, without applying any on-shell approximation. So far, this approach has been applied to loop-induced boson-pair production, and its gauge dependence in general multi-boson processes is still to be verified.

An important aspect when dealing with polarisation analyses is the reference-frame dependence of polarised-boson signals, motivated by the fact that polarisation vectors are not Lorentz-covariant quantities.
While different reference-frame choices have been made for Run-2 analyses, there are some which are better motivated than others. In this article, the centre-of-mass (CM) frame of the boson pair is employed to define polarisations. This choice is the most used in LHC di-boson analyses \cite{Aaboud:2019gxl,CMS:2021icx,ATLAS:2022oge,ATLAS:2023zrv,ATLAS:2024qbd}, and is well motivated from a theoretical viewpoint owing to the back-to-back kinematics of the bosons \cite{Denner:2020bcz} and the stability of polarisation fractions against higher-order corrections \cite{Denner:2021csi,Le:2022lrp}.

\subsection{Input parameters and event selection}\label{subsec:setupspec}
In this section we detail the SM parameters and fiducial selections considered for all presented numerical calculations.

The on-shell masses and widths of weak bosons are taken from \citere{Workman:2022ynf}:
\begin{eqnarray}\label{eq:weakparOS}
  \MWOS   =&  80.377  \GeV \,,\quad
  \GWOS   =&  2.085   \GeV \,,\quad\nnb\\
  \MZOS   =&  91.1876 \GeV \,,\quad
  \GZOS   =&  2.4952  \GeV \,,
\end{eqnarray}
and converted into pole values through the relations \cite{Bardin:1988xt},
\begin{eqnarray}\label{eq:weakCONV}
  M_V =\frac{ \MVOS }{ \sqrt{1+{(\GVOS/\MVOS)}^2 }}\,,\quad \Gamma_V =\frac{ \GVOS }{ \sqrt{1+{(\GVOS/\MVOS)}^2 }}\,,\quad V=\PW,\,\PZ\,.
\end{eqnarray}
In NWA calculations the partial decay width for $\PZ\rightarrow \Pe^+\Pe^-(\mu^+\mu^-)$ boson is computed at LO when providing LO and NLO QCD predictions, while it is computed
at NLO EW when providing NLO EW predictions. The branching is normalised to the Z-boson pole width that is used as an input to the Monte Carlo.

The electroweak coupling is evaluated in the $G_\mu$ scheme,
\beq\label{eq:alphaGmu}
\alpha_{G_\mu}\,=\,\frac{G_{\rm F}\sqrt{2}}{\pi}\MW^2\left(1-\frac{\MW^2}{\MZ^2}\right)\,,\qquad G_{\rm F}=1.16638\cdot 10^{-5}\GeV^{-2}\,.
\eeq
For the full off-shell calculations the complex mass scheme \cite{Denner:2005fg, Denner:2006ic, Denner:2019vbn} is used. 
In this scheme, the $\PW$ and $\PZ$ masses are replaced by the complex masses $\hat{M}_V^2 = M_V^2 - i M_V \Gamma_V$ 
($V=\PW,\PZ$) in the $S$-matrix elements, including the weak mixing angle $c_W^2 = \hat{M}_{\PW}^2/\hat{M}_{\PZ}^2$. 
However, the EW coupling $\alpha_{G_\mu}$ is kept real and calculated from the real parts as in Eq.~\ref{eq:alphaGmu}. 
For the DPA calculations, the real pole values $M_V$ are used everywhere, and the widths enter only in the denominator 
of the two resonances.

The top-quark and Higgs boson enter NLO EW and gluon-initiated loop-induced corrections. Their masses are taken from \citere{Workman:2022ynf}, while their widths are set to zero\footnote{Finite-width effects for the top quark and the Higgs boson are negligible because top quarks only appear in loops and the Higgs boson is always far off-shell due to the kinematical selections employed in the presented studies.},
\begin{eqnarray}\label{eq:topHiggsMasses}
  \Mt   =&  172.69\GeV \,,\quad
  \MH   =&  125.25\GeV \,.
\end{eqnarray}
Massless leptons, a unit CKM matrix and the five-flavour scheme are assumed.
The \sloppy \textsc{NNPDF31\_nnlo\_as\_0118\_luxqed} set \cite{Ball:2017nwa,Bertone:2017bme} is used as a default. This PDF set is accessible through the LHAPDF interface \cite{Buckley:2014ana} with id. 325100.
The $\overline{\rm MS}$ factorisation scheme for initial-state collinear singularities is understood for both QCD and EW corrections.
The running of the strong coupling $\alphas$ is extracted from the PDF set (\eg through the LHAPDF interface). The central renormalisation and factorisation scales ($\mu_{\rm R}$ and $\mu_{\rm F}$) are set to the $\PZ$-boson pole mass,
\beq
\mu_0=\MZ\,.
\eeq
The QCD-scale uncertainties are estimated with 7-point scale variations of $\mu_0$,
\beq\label{eq:7p_var}
\left(\frac{\mu_{\rm R}}{\mu_0},\frac{\mu_{\rm F}}{\mu_0}\right) = 
(1/2,1/2), (1/2,1),(1,1/2),(1,1)(1,2),(2,1),(2,2)\,. 
\eeq

The setup of the most recent ATLAS Run-2 measurement
\cite{ATLAS:2023zrv} is employed for the simulations.
Photons are recombined with charged leptons and quarks with either a cone dressing or the Cambridge/Aachen algorithm \cite{Dokshitzer:1997in,Wobisch:1998wt} and resolution radius $R=0.1$.
The QCD jets are clustered with the anti-$k_{\rm t}$ algorithm \cite{Cacciari:2008gp} and resolution radius $R=0.4$.
In predictions matched to both QED and QCD parton showers, lepton dressing is performed first. Then, the remaining photons and QCD partons are clustered into jets. An electron--positron pair and a muon-antimuon pair are required to satisfy the following fiducial cuts \cite{ATLAS:2023zrv}:
\beqn\label{eq:fiducialvolume}
&  \pt{\Pe^\pm}>7\GeV\,,\quad |y_{\Pe^\pm}|<2.47\,,\quad \pt{\mu^\pm}>5\GeV\,,\quad |y_{\mu^\pm}|<2.7\,,&\nnb\\
&  \pt{\ell_{1(2)}}>20\GeV\,,{\textrm{ with }} \ell_{1(2)}= {\textrm{the (second) hardest-$p_{\rm T}$ lepton}}\,,& \nnb\\
&  {\rm \Delta} R_{\ell\ell'}>0.05\,, {\textrm{ with }} \ell,\ell'=\Pe^\pm,\mu^\pm\,, &\nnb\\
&  81\GeV < M_{\ell^+\ell^-} <101\GeV\,, {\textrm{ with }} \ell=\Pe,\mu\,,&\nnb\\
&  M_{\rm 4\ell}>180\GeV\,.&
\eeqn
The di-lepton and four-lepton invariant-mass cuts ensure that the production of two on-shell bosons is kinematically allowed.

\subsection{Monte Carlo tools}\label{subsec:mctools}
In this work, a number of Monte Carlo (MC) codes have been compared, both at fixed order and matched to PS.
We detail in the following the main features of the various codes, with a special focus on the way they
carry out the polarised-signal selection and simulation.\\

{\red \moca}\, (MOnte CArlo at NLO accuracy) is an in-house, multi-purpose MC-integration program,
which has been used for NLO-accurate polarised-boson calculations
in both inclusive di-boson production
\cite{Denner:2020bcz,Denner:2020eck,Denner:2021csi,Denner:2022riz,Denner:2023ehn}
and in vector-boson scattering \cite{Denner:2024tlu}.
It is interfaced to the most recent release of the {\scshape Recola}-1
tree-level and one-loop amplitude provider \cite{Actis:2012qn, Actis:2016mpe}
and to the {\scshape Collier} library for one-loop tensor reduction and integration \cite{Denner:2016kdg}.
\moca\, is capable of computing complete NLO corrections (of both EW and QCD type) to generic LHC processes,
full off-shell effects and complete spin correlations both at LO and at NLO accuracy.
The dipole formalism \cite{Catani:1996vz, Dittmaier:1999mb,Catani:2002hc} is used to subtract both QCD and QED IR singularities.
For polarised processes the DPA \cite{Denner:2000bj,Denner:2005fg,Denner:2019vbn} is employed
for all contributions to NLO cross sections
(Born, virtual, real, local and integrated subtraction counterterms), ensuring that all matrix elements
are evaluated in the same Lorentz reference frame.
For more details we refer to \citere{Denner:2020bcz,Denner:2021csi,Denner:2024tlu}.\\

{\red \stri} (SecToR Improved Phase sPacE for real Radiation) is a \textsc{c++} implementation of the four-dimensional formulation of the sector-improved residue subtraction scheme \cite{Czakon:2010td, Czakon:2014oma, Czakon:2019tmo} which automates the subtraction and the numerical MC integration through NNLO QCD. The framework was extended to support intermediate polarisations for EW bosons using DPA and NWA and has been used for several polarisation studies \cite{Poncelet:2021jmj,Pellen:2021vpi, Pellen:2022fom}.
Matrix elements are taken from external libraries or are implemented explicitly. Tree-level matrix elements for the Born, single and double real radiation contributions are taken from the \textsc{AvH} library \cite{Bury:2015dla}. The necessary one-loop amplitudes are taken from \OpenLoops 2 \cite{Buccioni:2019sur, Cascioli:2011va, Buccioni:2017yxi}. The two-loop amplitudes are implemented with the help of the \textsc{VVamp} project \cite{Gehrmann:2015ora}.
To model intermediate polarised bosons, the DPA is implemented following the conventions in Ref.~\cite{Denner:2021csi} for the on-shell projections and polarisation-vector definitions. Several checks have been performed on the integrated cross-section level and per phase space point. The total cross section for off-shell $\PZ\PZ$ production was checked against \textsc{Matrix} at NNLO QCD \cite{Grazzini:2017mhc}. The polarised one-loop amplitudes obtained from a modified version of OpenLoops 2 were checked at the amplitude level against the private version of \textsc{Recola}-1 used in \cite{Denner:2020bcz} for various DPA setups.\\

{\red \mulb} (Multi-Boson production) is a private MC computer program to calculate polarised cross sections for 
multi-boson production processes in the DPA approach. The current version of the program can perform NLO QCD+EW calculations for $\PZ\PZ$, $\PW^\pm \PZ$, 
and $\PW^+\PW^-$ processes. Results for the $\PW\PZ$ and $\PW^+\PW^-$ cases have been published in \cite{Le:2022lrp,Le:2022ppa,Dao:2023kwc,Dao:2024ffg}.
The ingredients of this program include the helicity amplitudes for the production and decay processes, generated by 
{\sc FeynArts} \cite{Hahn:2000kx} and {\sc FormCalc} \cite{Hahn:1998yk}, and the in-house library {\sc LoopInts} for one-loop integrals. 
The tensor one-loop integrals are calculated using Passarino-Veltman reduction \cite{Passarino:1978jh}, 
while the scalar integrals are computed as in \cite{'tHooft:1978xw, Nhung:2009pm, Denner:2010tr}.  
The phase space integration is done using the MC integrator {\sc BASES} \cite{Kawabata:1995th}, with the help of resonance mapping routines publicly available in {\sc VBFNLO} \cite{Baglio:2024gyp}. 
The code has been carefully checked by making sure that all UV and IR divergences cancel and singular limits of the dipole 
subtraction terms behave correctly. 
For details the reader is referred to \cite{Le:2022ppa,Dao:2023kwc}.\\

{\red BBMC} (Boson-Boson Monte Carlo) is a general-purpose in-house MC integration code that uses {\sc Recola}-1 \cite{Actis:2012qn, Actis:2016mpe} as an amplitude provider and the {\sc Collier} library \cite{Denner:2016kdg} to compute one-loop scalar and tensor integrals. BBMC can compute integrated and differential cross-sections for arbitrary LHC processes at NLO QCD and NLO EW accuracy. Specifically it has been used for the calculation of polarised cross sections in diboson production \cite{Denner:2022riz,Denner:2023ehn} and vector-boson scattering \cite{Denner:2024tlu}.
The IR singularities are treated with the dipole subtraction scheme \cite{Catani:1996vz, Dittmaier:1999mb}. For the computation of polarised cross-sections at NLO accuracy, BBMC relies on the DPA \cite{Denner:2019vbn}. 
At NLO EW, the IR divergences associated to photon emission off intermediate $\PW$ bosons are treated in the DPA with an extension of the massive dipole subtraction scheme \cite{Dittmaier:1999mb,Catani:2002hc} and with a new counterterm tailored to the decay of $\PW$ bosons \cite{Denner:2023ehn,Denner:2024tlu}. 
The DPA and polarisation-selection implementation is analogous 
to the one in \moca. More details can be found in \citeres{Denner:2020bcz,Denner:2021csi,Denner:2024tlu}.\\

{\red \pwhg} (POsitive Weight Hardest Emission Generator)  is a general-purpose MC framework \cite{Alioli:2010xd} aimed at NLO-accurate
calculations matched to PS programs following the multiplicative {\scshape PowHeg} scheme
\cite{Nason:2004rx,Frixione:2007vw}.
The specific package used for this work \cite{Pelliccioli:2023zpd}
is based on a previous implementation of
di-boson processes \cite{Chiesa:2020ttl} in the {\scshape Res} version \cite{Jezo:2015aia},
which is capable of treating
radiative emissions off resonance propagators and decay products.
The code can compute any singly or doubly polarised di-boson process ($\PW\PW,\,\PW\PZ$ and
$\PZ\PZ$) in the fully leptonic decay channel at NLO QCD accuracy matched to PS, including hadronisation and multi-parton interactions (MPI). Di-boson processes can also be computed with unpolarised bosons or including complete off-shell effects.
Similarly to \moca~and \bbmc, this \pwhg~package is based on an interface to the {\scshape Recola}-1 
\cite{Actis:2012qn, Actis:2016mpe} and {\scshape Collier} libraries \cite{Denner:2016kdg}.
The sector subtraction scheme \cite{Frixione:1995ms} is used to subtract QCD IR singularities of initial-state kind (ISR). For (un)polarised processes,
the DPA \cite{Denner:2000bj,Denner:2005fg,Denner:2019vbn}
is used throughout the calculation of Born, virtual and real contributions.
The technical details of the \pwhg~realisation of the DPA are shown
in Ref.~\cite{Denner:2020bcz,Denner:2021csi,Denner:2024tlu}.
For the PS-matched predictions, NLO QCD corrections have been matched to {\scshape Pythia 8.244} \cite{Sjostrand:2014zea,Sjostrand:2019zhc} QCD and QED showers. Since only QCD corrections are included in the hard process, the QCD PS starting scale is given by the transverse momentum of the QCD ISR, while the QED PS has the partonic centre-of-mass energy as starting scale for ISR and the boson mass, the default in {\scshape Pythia}, for radiation off leptons. Hadronisation effects and MPIs can be enabled on top of the PS-matched simulations.\\

{\red \sher} (Simulation of High-Energy Reactions of PArticles) is a general-purpose MC event generator capable of simulating fully realistic particle collision events for arbitrary processes up to NLO QCD and approximate NLO EW~\cite{Sherpa:2024mfk}, including the simulation of the PS~\cite{Schumann:2007mg}, QED radiation~\cite{Schonherr:2008av}, hadronisation~\cite{Chahal:2022rid}, and MPIs. Polarised cross sections can be calculated for all LO tree-level processes involving intermediate vector bosons~\cite{Hoppe:2023uux} by employing an implemented NWA~\cite{Hoche:2014kca}. The vector-boson production and decay are modelled with a spin-correlation algorithm~\cite{Knowles:1987cu, Knowles:1988vs, Collins:1987cp, Richardson:2001df}. Off-shell effects are partially retained through a mass-smearing algorithm.
\sher~computes the complete helicity-dependent amplitude for the intermediate EW bosons, on top of an otherwise unpolarised simulation run. All polarisation contributions, including interference terms, are output as additional event weights. 
\sher~can provide fixed-order LO as well as PS-matched polarised predictions up to approximate NLO QCD (nLO QCD). Furthermore, multi-jet merging is available at both LO and nLO. Currently, nLO QCD corrections can only be included at production level. 
Tree-level polarised matrix elements are provided by the built-in matrix element generator {\sc Comix}~\cite{Gleisberg:2008fv}, while loop matrix elements are supplied by \OpenLoops~\cite{Cascioli:2011va, Buccioni:2019sur} except for loop-induced processes. For loop-induced processes, the EW-boson production is simulated based on {\sc Recola}-1 amplitudes \cite{Actis:2016mpe} mapped to the \Sherpa helicity-vector basis \cite{Hoppe:2025abc}. 
PS matching in \Sherpa is performed using its MC@NLO variant~\cite{Hoeche:2011fd}. 
The polarisation fractions are calculated based on different amplitudes, depending on the event type: for hard events ($\mathbb{H}$) and resolved standard events ($\mathbb{S}$), real-emission amplitudes are used, while the polarisation fractions for unresolved $\mathbb{S}$ events rely on Born-level amplitudes only, leading to approximate NLO accuracy in QCD.
For this study, PS-matched results are produced in three different modes: QCD shower only, QCD shower+QED radiation, and hadron-level event generation including also hadronisation, beam remnants and MPI effects, both at LO and n(N)LO\footnote{Only polarised calculations in \sher~require the described approximation. Both the unpolarised on- and off-shell results are fully NLO accurate.}. Photon splitting to fermions is disabled.
Multi-leg merging is based on the CKKW algorithm \cite{Lonnblad:2001iq, Hoeche:2009rj, Gehrmann:2012yg, Hoeche:2012yf, Hoeche:2014rya, Hoche:2010kg}. In this study a LO merged sample with up to two additional jets as well as an n(N)LO merged setup with 0,1 jet@n(N)LO and 2,3jet@LO are provided. Both samples assume a merging scale of $20\GeV$ and include all realistic event generation effects except photon splitting. In any case, no generator level selection is used except for a mass window of 66~GeV$<M_{\ell^+\ell^-}<$116~GeV for the off-shell calculations. \\

{\red\madg}\ ({\madgraph 5} a Monte Carlo at NLO) is an automated, general-purpose, MC simulation framework capable of modelling hard-scattering processes 
up to NLO in both QCD and EW couplings~\cite{Stelzer:1994ta,Alwall:2014hca,Frederix:2018nkq}.
It is used heavily by experimental collaborations
as well as the theory community for studying both SM processes and new phenomena.
For this study \madg~(v2.9.18) is used to produce two types of polarised and unpolarised samples:
the tree-level process $q\bar{q} \rightarrow\PZ\PZ \to 4\ell$ at $\mathcal{O}(\alpha^4)$, and the loop-induced process $\Pg\Pg \rightarrow\PZ\PZ \to 4\ell$ at $\mathcal{O}(\alpha^4\alphas^2)$.
For tree-level processes, the LO simulation of 
polarised or unpolarised intermediate resonances \cite{BuarqueFranzosi:2019boy} relies either on a diagram selection (BW) or on a spin-correlated NWA~\cite{Artoisenet:2012st}. 
The polarisations and spin correlations of intermediate EW bosons, which can be defined in arbitrary reference frames, are modelled through the decomposition of bosonic 
propagators into their transverse, longitudinal, and auxiliary components.
Gauge-dependent auxiliary polarisation vectors are 
needed to preserve gauge invariance in the case of off shell EW bosons but necessarily vanish in the on-shell limit.
%
For loop-induced processes, weak-boson polarisation modes are singled out at the level of amplitudes by directly modifying Feynman rules of propagators \cite{Javurkova:2024bwa}.
Unpolarised, intermediate vector-boson fields are decomposed into sums of helicity-polarised states, each with a separate propagator.
Spin correlations are preserved by generating Feynman diagrams  
for the full off-shell process and keeping only diagrams with two $s$-channel $\PZ$ bosons attached to a quark loop. 
Off-shell effects for polarised weak bosons are modelled using a Breit-Wigner modulation.
For PS-merged simulations, polarised-boson pairs have been simulated with up to 1(2) additional QCD partons for the loop-induced (tree-level) processes, according to the MLM scheme~\cite{Mangano:2006rw,Alwall:2007fs}
as implemented in {\sc Pythia8.237} \cite{Sjostrand:2019zhc}, with a minimum transverse momentum of $15\GeV$ on additional partons.
Hadronisation, MPI and QED showering are disabled. The ATLAS A14 central tune and the NNPDF2.3LO PDF set are used.
For loop-induced contributions, a diagram filter ensures that only genuine loop-induced
diagrams are included at orders $\order{\alphas^2\alpha^4}$ and $\order{\alphas^3\alpha^4}$.

\section{Results}\label{sec:results}
In this section we present all numerical results obtained with the MC codes and event generators described in \refse{subsec:mctools}. A broad validation has been carried out at LO and at NLO (see \refse{sec:fo_results}), while both fixed-order and PS-matched/merged results are considered for the phenomenological analysis of \refse{sec:ps_results}, \ref{sec:fractions} and \ref{sec:comp_exp}. On top of integrated fiducial cross sections, we will consider differential distributions, focusing on both angular and transverse-momentum observables.

\subsection{Fixed-order validation}\label{sec:fo_results}
Before detailing the results of the comparison, we stress again that the various MC codes considered in this work rely on different approaches to define polarised signals for intermediate $\PZ$ bosons. On the one hand, \moca, \stri, \mulb, \bbmc, and \pwhg~rely on similar realisations of the DPA. On the other hand, \sher~and \madg~rely on the NWA and BW approaches, respectively. As already mentioned in \refse{subsec:setup}, the \madg~predictions (BW) do not rely on any on-shell approximation but rather on simple non-resonant diagram removal, leading to a potential gauge dependence of the results \cite{Javurkova:2024bwa}.
For completeness we present results not only for the unpolarised and polarised $\PZ\PZ$ production in the various code realisations, but also the results in the full off-shell picture. This is relevant to assess the impact of off-shell effects, or in other words of the non-resonant background which enter polarised-template fits of LHC data. 

The fiducial cross sections in the ATLAS setup described in \refse{subsec:setupspec} are reported at the available perturbative orders in Table~\ref{tab:LOfid}.
All codes are capable of computing polarised and unpolarised cross sections at tree level in the quark-induced production channel, namely at order $\mc O(\alpha^4)$. The numerical values are shown in the upper block of Table~\ref{tab:LOfid}.
\begin{table}[ht]
  \begin{center}
\hspace*{-0.4cm}    \begin{tabular}{lllllll}
      \hline\rule{0ex}{2.7ex}
      \cellcolor{blue!9} code & \cellcolor{blue!9} full & \cellcolor{blue!9} unpol. & \cellcolor{blue!9} LL & \cellcolor{blue!9} LT & \cellcolor{blue!9} TL & \cellcolor{blue!9} TT \\
      \hline\\[-0.4cm]
      \multicolumn{7}{c}{\cellcolor{yellow!9} Tree level ($q\bar{q}$)} \\
      \hline\\[-0.25cm] 
      \moca &  $11.336(1) $   & $ 11.242(1)     $ & $ 0.6574   (1 ) $ & $ 1.3332   (2 ) $ & $ 1.3370   (2 ) $  & $ 7.7874   (8 ) $  \\ 
      \stri &  $11.3357(4)$   & $ 11.2451   (2 )$ & $ 0.6560   (0 ) $ & $ 1.3326   (0 ) $ &  $ 1.3365   (0 ) $ & $ 7.7925   (1 ) $  \\
      \mulb &  $11.3363(2)$  & $ 11.2393   (3 )$ & $ 0.6572   (0 ) $ & $ 1.3329   (1 ) $ & $ 1.3366   (1 ) $  &  $ 7.7846   (2 ) $  \\
      \bbmc &  $11.3372(4)$   & $ 11.2424   (3 )$ & $ 0.6574   (0 ) $ & $ 1.3333   (1 ) $ &  $ 1.3372   (1 ) $ &  $ 7.7872   (2 ) $ \\
      \pwhg &  $11.335(1)$   & $ 11.245(1)$ & $ 0.6575(1)$ &  $ 1.3333(1)$ & $ 1.3374(1)$ &$ 7.7885(8)$ \\    
      \sher~(NWA) &  $11.337(4) $   & $ 11.513(4)     $ & $ 0.6767(4) $     & $ 1.3538(6) $ & $ 1.3734(6) $  & $ 7.952(3) $  \\
      \madg~(BW) &  $11.38(2)  $   & $ 11.29(2)      $ & $ 0.660(1)  $     & $ 1.335(2)  $ & $ 1.338(2) $  &  $ 7.81(1)  $ \\
      \hline\\[-0.4cm]
      \multicolumn{7}{c}{\cellcolor{yellow!9} Loop induced ($\Pg\Pg$)} \\
      \hline\\[-0.25cm]
      \moca     &  $ 1.6968   (6 )$  & $ 1.6978   (6 ) $   & $ 0.0914  (0 )  $  &  $ 0.0360   (0 ) $  & $ 0.0356   (0 ) $   & $ 1.5360   (5 ) $  \\ 
      \stri     &  $ 1.682   (7 ) $  & $ 1.700     (2 )$   & $ 0.0912   (1 ) $  &  $ 0.0360   (0 ) $  & $ 0.0357   (0 ) $   & $ 1.538   (2 ) $ \\ 
      \mulb     &  $  - $            & $ 1.6981   (9 ) $   & $ 0.0913   (1 ) $  &  $ 0.0360   (0 ) $  & $ 0.0357   (0 ) $   & $ 1.5363   (8 ) $ \\
       \sher~(NWA) &  $1.6967(5) $   & $ 1.7352(5)     $ & $  0.09371(5)	 $     & $ 0.03679(1)	 $ & $0.03644(1)	 $  & $ 1.5696(4) $  \\
       \madg$^{*}$(BW)  &  $ 1.699   (6 ) $  &  $ 1.697    (6) $   & $ 0.0902   (3 ) $  &   $ 0.0355   (1 ) $ & $ 0.0359   (1 ) $   & $ 1.539(6) $ \\ 
            \hline\\[-0.4cm]
      \multicolumn{7}{c}{\cellcolor{yellow!9} NLO QCD} \\
      \hline\\[-0.25cm]
      \moca  &  $ 15.282(1)$   & $15.158(2)$   & $0.8899(3)$  & $ 1.9313(5)$ & $1.9243(2)$ &  $10.2095(9)$ \\
      \stri  &  $15.284(3)$   & $ 15.159(1) $  & $0.8899(1)$  &  $ 1.9305(1)$ & $ 1.9241(1) $ & $ 10.2098(7) $ \\
      \mulb  &  $ - $   & $15.1575(9) $  & $ 0.88997(6) $  & $1.9305(1)$ & $ 1.9240(1) $ &  $ 10.2106(6) $ \\
      \bbmc  &  $15.284(1)$   & $15.158(1)$  & $0.8898(1)$  &  $1.9306(2)$ & $1.9240(2)$  & $10.2085(7)$ \\
      \pwhg$^{1}$  &  $15.280(2)$   & $15.156(2)$ & $ 0.8909(2)$ & $ 1.9306(4) $ & $ 1.9239(5)$ & $ 10.206(1) $\\    
      \pwhg$^{2}$  &  $15.330(7)$   & $15.177(7)$ & $0.8918(4 )$ & $ 1.9366(9 ) $ & $ 1.9291(9)$ & $ 10.215(5 )$\\    
      \sher$^{\dag}$~(NWA) &  $15.304(4)$   & $15.441(5)$ & $0.9266(5)$  & $2.093(1)$    & $2.041(1)$     & $10.289(4)$ \\	
      \hline\\[-0.4cm]
      \multicolumn{7}{c}{\cellcolor{yellow!9} NNLO QCD} \\
      \hline\\[-0.25cm]
      \stri$^{\ddag}$ &  $ 16.19(2) $   & $16.06(1)$   & $0.9756(9)$  & $ 2.107(2)$ & $2.094(2)$ &  $10.63(1)$ \\
      \hline\\[-0.4cm]
      \multicolumn{7}{c}{\cellcolor{yellow!9} NLO EW} \\
      \hline\\[-0.25cm]
      \moca  &  $ 10.080(2)$   & $10.0213(8)$  & $0.59068(9)$  & $1.1994(1)$ & $1.20293(9)$ & $6.9129(3)$  \\
      \mulb  &  $ - $   & $10.0203(3)$  & $ 0.59058(2) $  & $1.19926(4)$  & $1.20294(4)$ & $6.9121(3)$  \\
      \bbmc  &  $ 10.082(2)$   &  $10.0203(4)$ &  $ 0.59057(4)$ & $1.19949(6)$ & $1.20308(9)$ & $ 6.9125(3) $  \\
      \hline
    \end{tabular}\qquad
  \end{center}
  \caption{Fiducial cross sections in $\fb$ for off-shell, unpolarised and doubly polarised $\PZ\PZ$ production and decay in the LHC setup described in Eq.~\ref{eq:fiducialvolume}. The results are for the DPA if not stated otherwise. $^{*}$While \madg~tree-level cross sections are obtained with the standard BW approach \cite{BuarqueFranzosi:2019boy}, the corresponding loop-induced contribution are obtained with the tailored UFO model proposed in \citere{Javurkova:2024bwa}.   
    $^{\dag}$\sher~results in the NWA \cite{Hoppe:2023uux} for polarised signals reach approximate NLO accuracy in QCD (nLO)
    and include partial resummation effects from the QCD-shower truncation after the first emission.
    $^{1}$ \pwhg~results feature exact NLO QCD accuracy at fixed order and $^{2}$
    additionally, including resummation effects coming from the Sudakov form factor (LHE level).
    $^{\ddag}$\stri~results at NNLO QCD do not include loop-induced $\Pg\Pg$ contributions.   
    \label{tab:LOfid}
  }
\end{table}
All results obtained with the DPA approach are compatible within numerical-integration uncertainties (shown in parentheses), for both unpolarised and polarised signals. The \madg~results  are also compatible with the DPA ones, and a permille-level discrepancy is found for the full off-shell calculation. The off-shell result from \sher~is also very close to the others, while the NWA values for both unpolarised and polarised processes are about 2-to-3\% higher than the corresponding DPA ones. As it was already observed with two different implementations of polarised signals in \stri~for $\PW^+\PW^-$ production \cite{Poncelet:2021jmj}, the \sher~NWA realisation overshoots the off-shell result by about $1.3\%$, while the BW strategy in \madg~gives a similar underestimate of the full off-shell cross section by less than $1\%$. In any case, the unpolarised BW, NWA and DPA results reproduce the off-shell ones with an accuracy of $1\%$, complying with the expected intrinsic accuracy of the two approximations \cite{Uhlemann:2008pm,Denner:2019vbn}.

\begin{figure}
  \centering
  \includegraphics[width=0.53\textwidth]{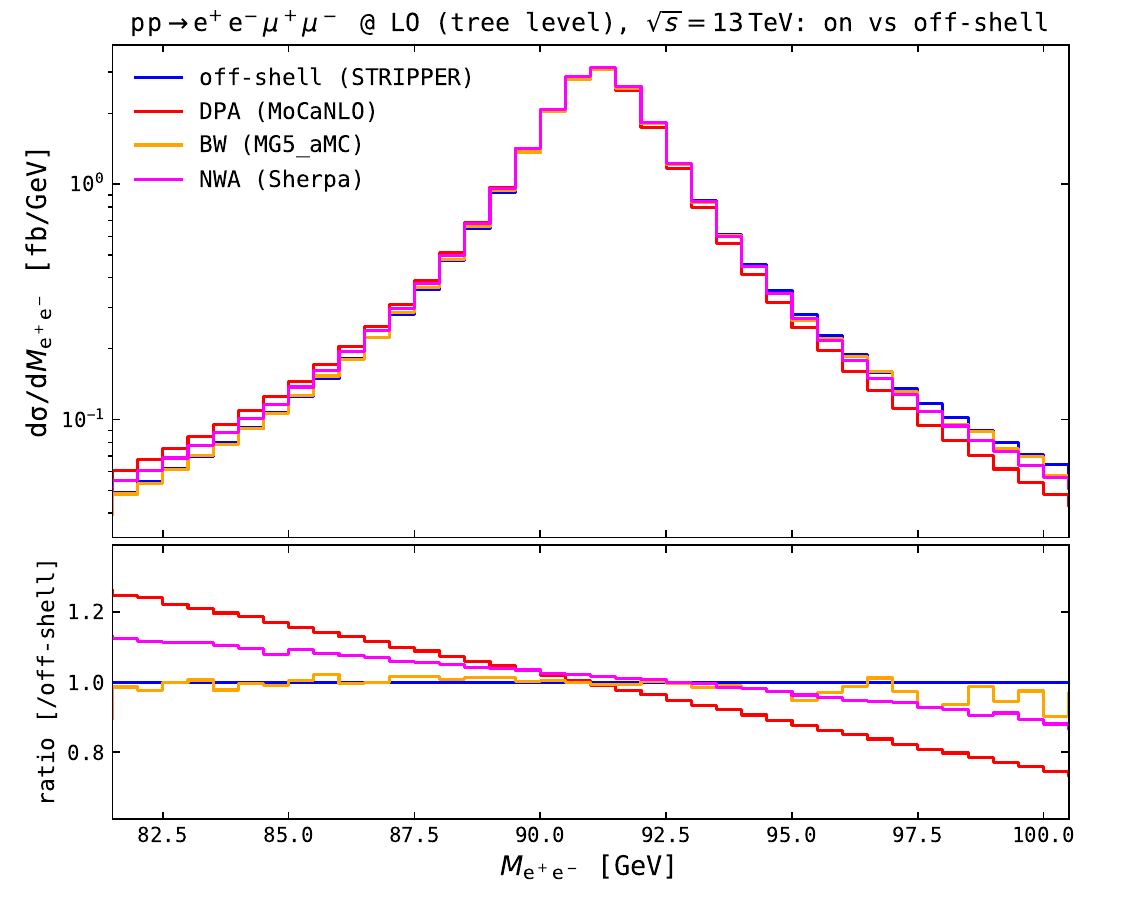}
  \caption{Differential distributions in the electron-positron invariant mass at tree level in the setup described in Eq.~\ref{eq:fiducialvolume}.
     Absolute distributions obtained with the BW, NWA and DPA approaches for the unpolarised process are compared to the full off-shell one in  the top panel. The ratios over the full off-shell result are shown in the bottom panel.  
  }
  \label{fig:onoffshell}
\end{figure}

The slight differences amongst the NWA, DPA and BW approaches are rather independent of the polarisation state (in fact, very similar polarisation-fraction predictions are obtained), and can be better understood looking at the unpolarised distributions in the invariant mass of one same-flavour charged-lepton pair, shown in \reffi{fig:onoffshell}.
The discrepancies, sizeably growing when going far from the $\PZ$-boson pole mass, can be traced back to the different off-shell smearing of resonances in \madg~and \sher~and the DPA calculations. In \madg, a standard decay-chain syntax was used \cite{BuarqueFranzosi:2019boy}, automatically preserving spin correlations between production and decay of the bosons, and therefore including finite-width effects via the Breit-Wigner modulation in the propagators.
In \sher, the mass-smearing algorithm is applied after the computation of the production and decay of on-shell vector bosons. The off-shell-ness of the boson is chosen according to the Breit-Wigner distribution. The final-state momenta are redistributed to account for the boson virtuality, preserving the flight direction in their joint CM frame. In DPA calculations, the off-shell-ness of intermediate bosons is automatically preserved both in propagator denominators and in the phase-space weights, while matrix-element weights (the amplitude numerator) are projected via on-shell mappings. This leads to a change in the weights for events above and below the pole mass.

\begin{figure}[t]
  \centering
  \subfigure[Positron--antimuon invariant mass   \label{fig:Mepmup}]{\includegraphics[width=0.49\textwidth, page=2]{fig/final_validation_plots_logg.pdf}}
  \subfigure[Positron--electron azimuthal-angle separation \label{fig:dphiee_gg}]{\includegraphics[width=0.49\textwidth, page=1]{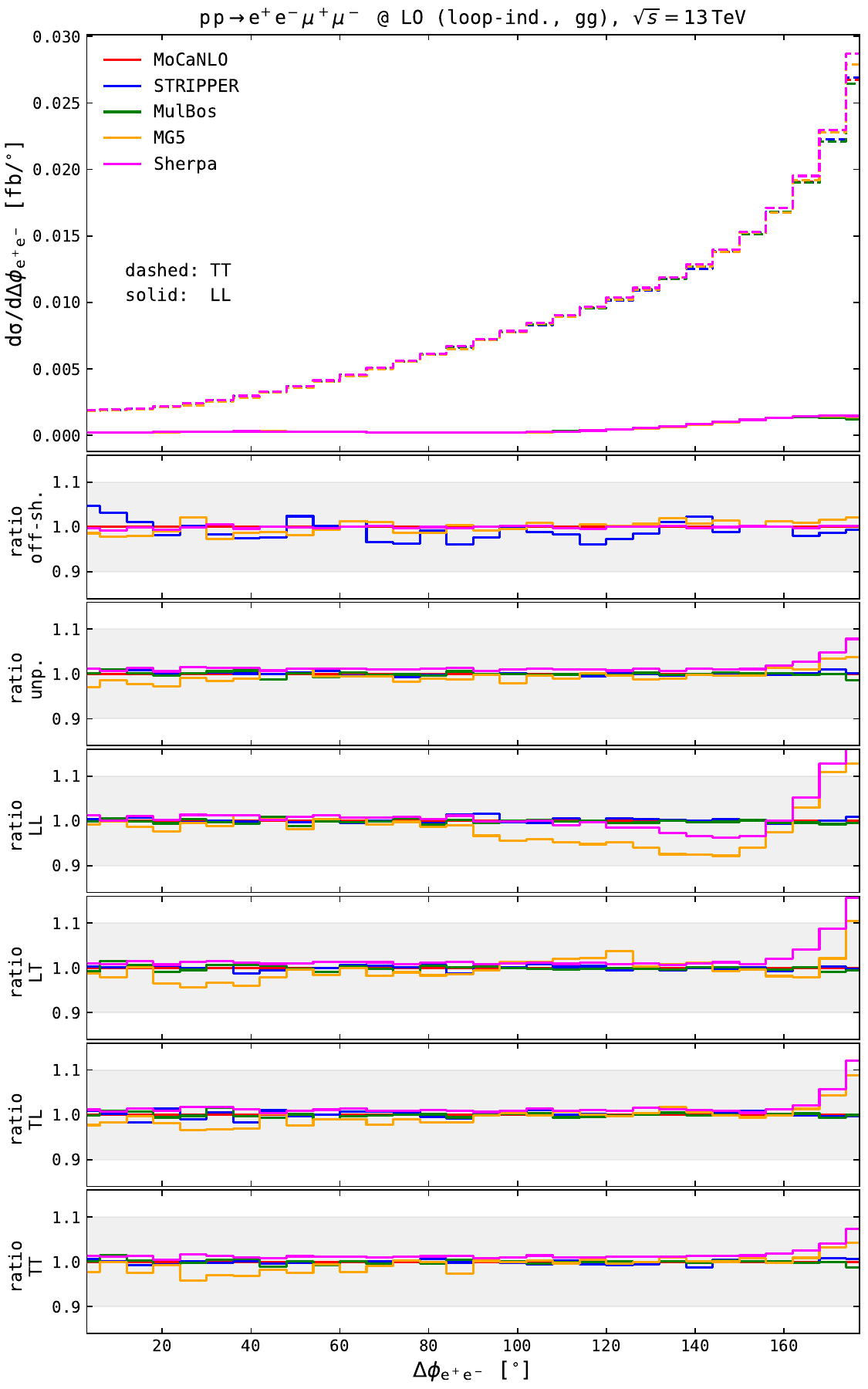}}
  \caption{
    Differential distributions for the loop-induced $\Pg\Pg$ contributions 
    in the setup described in Eq.~\ref{eq:fiducialvolume}, 
    obtained with various MC generators. 
    The invariant mass of the positron--antimuon pair (left) and the azimuthal-angle separation between the positron and the electron (right) are considered.
     Absolute distributions for the LL (solid curves) and TT (dashed curves) polarisation modes are shown in top panels. 
    Ratio plots for the various modes (off-shell, unpolarised, LL, LT, TL, TT) are shown in lower panels, w.r.t. \moca~results.
    Shaded gray bands in the lower panels span ratios between 0.85 and 1.15.
  }\label{fig:valid_gg}
\end{figure}

In the second block of Table~\ref{tab:LOfid} we show the results for the loop-induced production channel with two initial-state gluons. The \moca, \stri~and \mulb~results are again obtained with the same DPA approach as used for tree-level cross sections. The \sher~results are obtained in the NWA. The \madg~one is obtained with the novel strategy of defining polarised signals starting from a modification of Feynman rules \cite{Javurkova:2024bwa}. In this approach, the intermediate bosons are not set on-shell, therefore embedding automatically off-shell effects. This approach works well for inclusive $\PZ$-pair production, where non-resonant effects are almost absent. In more intricate processes with unsuppressed non-resonant topologies, it can lead to large deviations from the NWA and DPA approaches, owing to the gauge dependence of the resonant signals. For the process at hand, the agreement amongst the calculations is rather good at the integrated level, for all polarisation states as well as for the unpolarised and full off-shell modelling. 
A very good agreement is found amongst different DPA calculations (\moca, \stri, \mulb) also at the differential level, as shown in \reffi{fig:valid_gg}. In the figures, the absolute distributions for the LL and TT polarisation states are shown in the main panels, while ratio plots (relative to the \moca~results in the DPA) are shown for all polarised states, for the unpolarised and for the full off-shell distributions.
In the case of the distribution for the invariant-mass of the positron-antimuon system shown in \reffi{fig:Mepmup}, the \madg~results (obtained with the novel UFO \cite{Javurkova:2024bwa}) are in fair agreement with the DPA ones for all polarisation states and for unpolarised calculations as well. The \sher~NWA results agree with the DPA ones up to 3\% deviations in low-mass region, compatible in any case with the intrinsic uncertainty of the on-shell approximations.
On the contrary, the distributions in the azimuthal distance between the positron and the electron (see \reffi{fig:dphiee_gg}) highlight a clear discrepancy between the NWA/BW results and the DPA ones. A marked shape distortion is found in the most populated region, $\Delta\phi_{\Pe^+\Pe^-}\approx \pi$, especially for the LL signal.
The same pattern, with smaller size, is found for other polarisation states.
This azimuthal observable is strongly correlated to the invariant mass of the electron-positron pair, where a similar behaviour is found as the one of \reffi{fig:onoffshell}, suggesting that one motivation for this deviation comes from the way (partial) off-shell effects are retained in the BW approach, compared to the DPA. The different deviations found for the various polarisation modes suggest that in the NWA/BW approach of \citere{Javurkova:2024bwa} the off-shell effects are not distributed in the same manner for longitudinal and transverse modes.

\begin{figure}
  \centering
  \subfigure[Positron--electron azimuthal-angle separation \label{fig:dphiee_nloqcd}]{\includegraphics[width=0.49\textwidth, page=1]{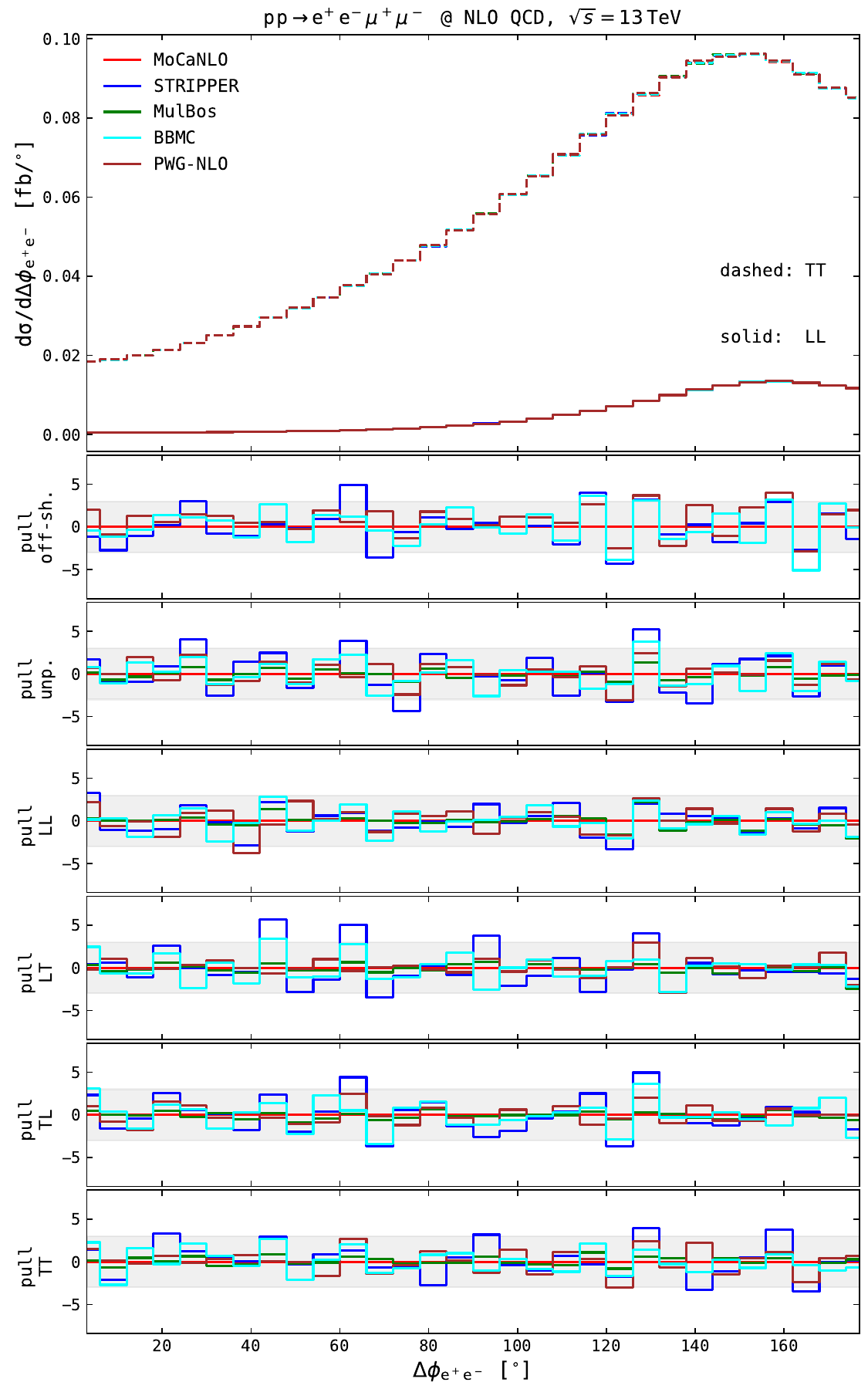}}
  \subfigure[Positron transverse momentum   \label{fig:ptep_nloqcd}]{\includegraphics[width=0.49\textwidth, page=2]{fig/final_validation_plots_nloqcd.pdf}}\\
  \caption{
    NLO QCD differential distributions 
    in the setup described in Eq.~\ref{eq:fiducialvolume}, 
    obtained with various MC generators. 
    The positron--electron azimuthal-angle separation (left) and the positron transverse momentum (right) are shown.
     Absolute distributions for the LL (solid curves) and TT (dashed curves) polarisation modes are shown in top panels. 
Pull curves for the various modes (off-shell, unpolarised, LL, LT, TL, TT) are shown in lower panels. 
    The pull is computed as the discrepancy w.r.t. \moca~results normalised to the quadrature-combined MC-integration uncertainty.
    Shaded gray bands in the lower panels span pulls between -3 and 3.   
  }\label{fig:validNLOQCD}
\end{figure}

The NLO QCD fiducial cross sections for the various polarisation states are presented in the third block of Table~\ref{tab:LOfid}. Differential results at this perturbative order are shown in \reffi{fig:validNLOQCD} for an angular and a transverse-momentum observable. 
The agreement amongst MC codes relying on the DPA is very good at both integrated and differential level. In both angular (see \reffi{fig:dphiee_nloqcd}) and transverse-momentum distributions (see \reffi{fig:ptep_nloqcd}), all codes generate results compatible within MC-integration uncertainties, up to statistical fluctuations in very few bins. In \reffi{fig:validNLOQCD} we have not shown LHE-level results from \pwhg~(unweighted events with the hardest real emission generated by \POWHEG~according to the Sudakov form factor), nor nLO results from \sher, while the corresponding fiducial cross sections are included in Table~\ref{tab:LOfid}. Such results are expected to deviate from fixed-order NLO-accurate ones, owing to the inclusion of resummed QCD effects at leading-logarithmic accuracy. A discussion of such effects is postponed to the end of this section.

The calculation of NNLO QCD corrections for polarised-boson pair production, so far carried out uniquely for $\PW^+\PW^-$ production in the two-charged-lepton decay channel \cite{Poncelet:2021jmj}, has been extended to the $\PZ\PZ$ case \cite{Poncelet:2025abc}. While a broad discussion of NNLO QCD results is given in \refse{sec:ps_results} as benchmarks for PS-matched and merged calculations, we point out here that NNLO QCD corrections are sizeable and different for the various polarisation states. Excluding $\Pg\Pg$ loop-induced contributions (which are formally of the same order $\mc O(\alphas^2\alpha^4)$ as genuine corrections to the tree-level di-boson process), the NNLO QCD corrections range between $+4\%$ for the TT state and $+10\%$ for the LL signal, relatively to the corresponding NLO QCD results. Such large corrections to the LL polarisation state are driven by large hard-real contribution, rather than virtual corrections.

\begin{figure}
  \centering
  \subfigure[Positron--antimuon azimuthal-angle separation \label{fig:dphiee_nloew}]{\includegraphics[width=0.49\textwidth, page=1]{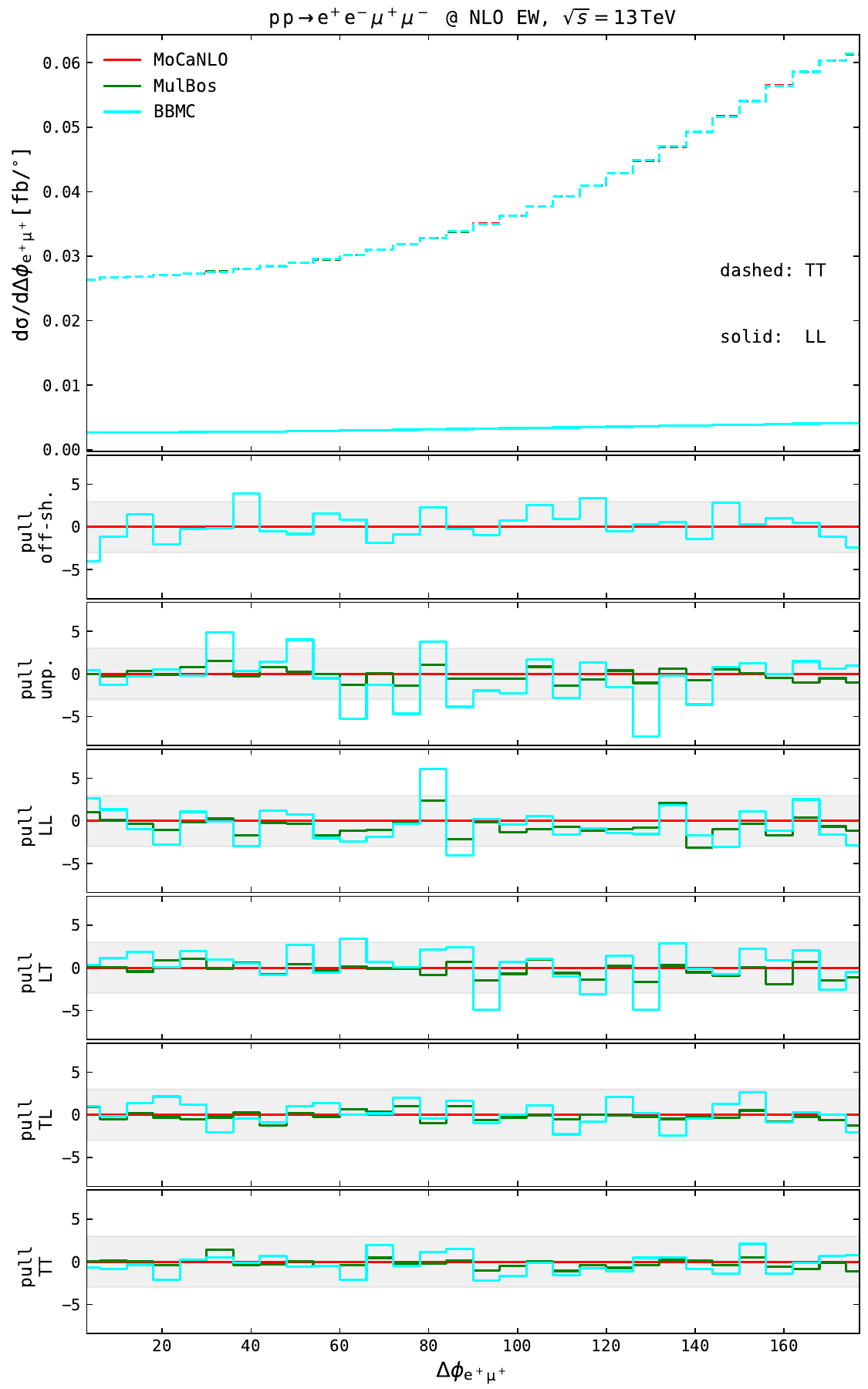}}
  \subfigure[Antimuon transverse momentum   \label{fig:ptep_nloew}]{\includegraphics[width=0.49\textwidth, page=2]{fig/final_validation_plots_nloew.pdf}}
  \caption{
    Differential distributions at NLO EW 
    in the setup described in Eq.~\ref{eq:fiducialvolume}, 
    obtained with various MC generators. 
    The positron--antimuon azimuthal-angle separation (left) and the antimuon transverse momentum (right) are considered.
    Same structure as \reffi{fig:validNLOQCD}.
  }\label{fig:valid_NLOEW}
\end{figure}

Three MC codes are capable of computing NLO EW corrections to polarised-boson pairs, all relying on the DPA approach \cite{Denner:2021csi,Le:2022lrp,Le:2022ppa,Denner:2023ehn,Dao:2023kwc}. As shown at integrated level (see last block of Table~\ref{tab:LOfid}), the agreement amongst the codes is almost perfect. Most of the distributions, like the angular one considered in \reffi{fig:dphiee_nloew}, highlight very good agreement amongst the MC codes at NLO EW even at differential level. Somewhat larger discrepancies are found in the most populated bins of some transverse-momentum distributions between \mulb~and {\sc Recola}-based tools (\bbmc~and \moca), as depicted in \reffi{fig:ptep_nloew}. 
These deviations, though significant in terms of integration uncertainties, are at the permille level and do not lead to visible shape distortions.

Additionally, they only affect polarisation states with at least one longitudinal boson, suggesting a bias in the \mulb~results which is enhanced for distributions which are suppressed in the tails. This bias is also present in the full off-shell LO distributions when comparing \mulb~to all the other tools (including \stri, \sher, \madg).

As observed in the main panel of \reffi{fig:ptep_nloew}, the LL distribution falls way faster than the TT one towards large transverse momenta. This behaviour is in agreement with the expectations of the Goldstone-boson equivalence theorem \cite{Cornwall:1974km,Vayonakis:1976vz,Chanowitz:1985hj,Gounaris:1986cr}, indicating a LO suppression of the LL signal by $1/s^2$ and $1/s$ with respect to the TT and mixed (LT, TL) signals, respectively \cite{Denner:2021csi}. The negative character of NLO EW corrections leads to an unphysically negative cross section at NLO EW for $\pt{\ell}\gtrsim 240\GeV$, implying the need for the inclusion of partial NNLO EW effects coming from squared one-loop amplitudes \cite{Bierweiler:2013dja,Denner:2021csi}.

We conclude this section by commenting on the resummation effects that are included in \pwhg~LHE-level and \sher~results.
\begin{figure}
  \centering
  \subfigure[Positron--electron azimuthal-angle separation \label{fig:dphiee_HOQCD}]{\includegraphics[width=0.49\textwidth, page=1]{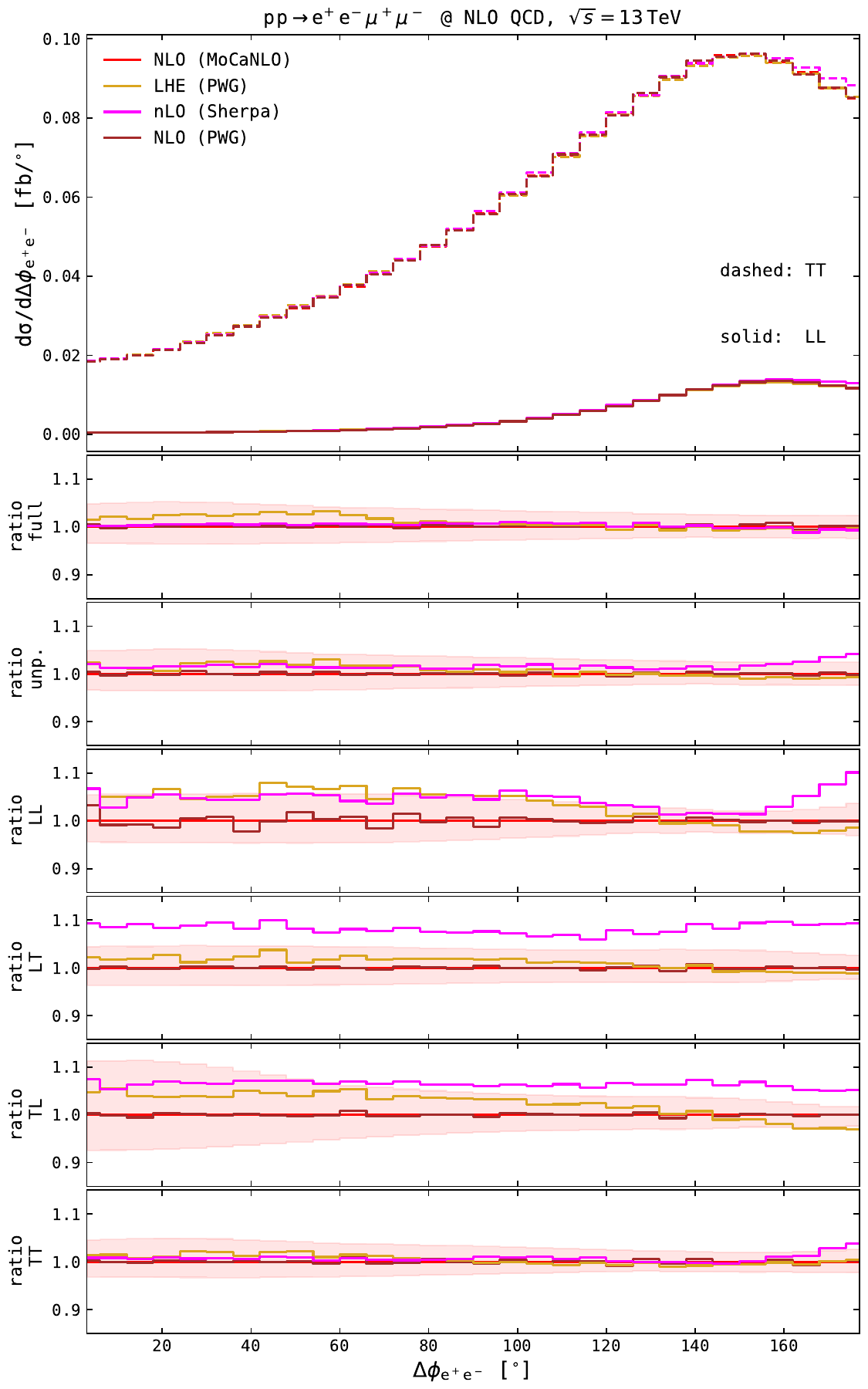}}
  \subfigure[Positron--electron-pair transverse momentum   \label{fig:ptee_HOQCD}]{\includegraphics[width=0.49\textwidth, page=2]{fig/final_validation_plots_resum.pdf}}\\
  \caption{
    Higher-order QCD effects on differential distributions 
    in the setup described in Eq.~\ref{eq:fiducialvolume}, obtained at 
    fixed order (NLO results from \moca~in red and \pwhg~in brown),
    and with resummation effects (LHE results with extact NLO accuracy from \pwhg~in goldenrod, 
    approximate NLO results from \sher~in magenta).
    The positron--electron azimuthal-angle separation (left) and transverse momentum (right) are shown.
    Absolute distributions for the LL (solid curves) and TT (dashed curves) polarisation modes are shown in top panels. 
    Ratios  w.r.t. to \moca~results for the various modes (off-shell, unpolarised, LL, LT, TL, TT) are shown in lower panels.
    Shaded bands in the lower panels represent QCD-scale uncertainties from MoCaNLO at NLO.    
  }\label{fig:validRESUM}
\end{figure}
The two MC codes are capable of producing unweighted events at exact (\pwhg) or approximate (\sher) NLO QCD accuracy, to be then matched to PS effects. Both in the additive {\sc MC@NLO} matching \cite{Frixione:2002ik,Hoeche:2011fd} used in \sher~and in the multiplicative matching \cite{Nason:2004rx} used in \pwhg, the leading-logarithmic approximation of the PS (which is accurate only at small transverse momenta) is improved to achieve NLO accuracy for inclusive observables. While both LHE-level \pwhg~results and \sher~results embed leading-logarithmic resummation regulating the Sudakov region at low transverse momentum of the $\PZ\PZ$ pair, only \pwhg~results feature exact NLO QCD accuracy for polarised and unpolarised signals. In the case of polarised signals, the \sher~results include approximate NLO QCD corrections, as only $\mathbb{H}$ events and resolved $\mathbb{S}$ events are computed with exact matrix elements. For unresolved $\mathbb{S}$ events, Born-level matrix elements are employed, neglecting one-loop corrections as well as unresolved soft and collinear emissions.  
The effect of this approximation has been proved to be small for doubly polarised $\PW\PZ$ inclusive production, by comparing with exact NLOPS predictions~\cite{Pelliccioli:2023zpd}.
Coming back to $\PZ\PZ$ fiducial cross sections in Table~\ref{tab:LOfid}, the fixed-order (NLO) results and those embedding resummation effects in \pwhg~(LHE) agree at the permille level, with tiny differences due to the fiducial cuts (the two cross sections would be identical in a fully inclusive setup). The nLO \sher~results for polarised states sizeably differ from the DPA results. For unpolarised and TT states, the 2\% deviation is within the intrinsic NWA accuracy, in line with LO findings. For other polarisation states, the discrepancy becomes larger (4\% for LL, 7\% for mixed states). Since resummation effects are not expected to give so sizable effects to the integrated fiducial cross sections, the results suggest that including exact virtual QCD corrections is important for longitudinal-boson signals. 
Resummation effects captured by the matching kernels in \pwhg~and \sher~can be appreciated mostly in the distribution-shape distortions.
In \reffi{fig:validRESUM}, the resummed distributions are compared to the fixed-order ones (\moca~and PWG-NLO distributions, already shown in \reffi{fig:validNLOQCD}) for two differential observables. The QCD-scale uncertainties (7-point scale variations) are shown in ratio plots as shaded red bands. In both angular observables (\reffi{fig:dphiee_HOQCD}) and transverse-momentum ones (\reffi{fig:ptee_HOQCD}), the \pwhg~and \sher~resummation kernels lead to shape distortions compared to fixed order, with largely different behaviours for the various polarisation states. While for the azimuthal distance between the positron and the electron, the LHE results from \pwhg~lie within QCD-scale uncertainty bands, the \pwhg~Sudakov form factor has slightly more evident effect on the $\PZ$-boson transverse-momentum distribution with the resummation damping the low-$p_{\rT}$ region and recovering the NLO result for moderate-to-large $p_{\rT}$. A similar situation is found in \sher~results, although with a different shape distortion compared to fixed order. For the LL state in particular, the \sher~resummation effects in the lowest bin of the $\PZ$-boson transverse momentum enhance the fixed NLO result by 13\%, while in the \pwhg~case they diminish it by 4\%. Towards larger transverse momenta, the \sher~curve approaches the fixed order faster than the LHE \pwhg~one does. This is a known feature of \pwhg~\cite{Alioli:2008tz,Melia:2011tj} $p_{\rT}$ distributions which are generally harder than those obtained with MC@NLO-like matching, owing to beyond-accuracy (NNLO) terms that become large in differential distributions. Though to a lesser extent, similar features are found in the TT distributions.

\begin{table}
  \begin{center}
    \begin{tabular}{lllll}
      \hline\rule{0ex}{2.7ex}
      \cellcolor{blue!9} & \cellcolor{blue!9}LL & \cellcolor{blue!9}LT & \cellcolor{blue!9}TL & \cellcolor{blue!9}TT \\
      \hline\\[-0.25cm]
NLO$_{\rm QCD}$    &   $ 0.8899 (3 ) ^{+ 3.1 \%}_{ -2.5 \%} $ &   $ 1.9313 (5 ) ^{+ 3.6 \%}_{ -2.9 \%} $ &   $ 1.9243 (2 ) ^{+ 3.6 \%}_{ -2.9 \%} $ &   $ 10.209 (1 ) ^{+ 2.8 \%}_{ -2.2 \%} $ \\ [0.1cm]
NNLO$_{\rm QCD}$   &   $ 0.976 (1 ) ^{+ 2.2 \%}_{ -1.9 \%} $ &   $ 2.107 (2 ) ^{+ 2.1 \%}_{ -1.9 \%} $ &   $ 2.094 (2 ) ^{+ 2.0 \%}_{ -1.8 \%} $ &   $ 10.63 (1 ) ^{+ 1.1 \%}_{ -1.0 \%} $ \\ [0.1cm]
NNLO$^{\rm (+)}$ &    $ 0.909 (1 ) ^{+ 2.9 \%}_{ -2.1 \%} $ &   $ 1.973 (2 ) ^{+ 2.7 \%}_{ -2.0 \%} $  &   $ 1.960 (2 ) ^{+ 2.6 \%}_{ -1.9 \%} $  &   $ 9.76 (1 ) ^{+ 1.6 \%}_{ -1.1 \%} $ \\[0.1cm]
NNLO$^{\rm (\times)}$ & $ 0.876(1) ^{ + 2.1  \%} _{ -1.9 \%} $ &   $ 1.895(2) ^{ + 2.0  \%} _{ -1.9 \%} $ &   $ 1.884(2) ^{ + 1.9  \%} _{ -1.8 \%} $ &   $ 9.439(9) ^{ + 0.9  \%} _{ -1.0 \%} $ \\[0.1cm]
\hline\\[-0.25cm]
NLOPS$_{\rm QCD}$  &  $ 0.8918 (3 ) ^{+ 3.0 \%}_{ -2.5 \%} $ &   $ 1.9367 (6 ) ^{+ 3.6 \%}_{ -2.9 \%} $ &   $ 1.9293 (6 ) ^{+ 3.5 \%}_{ -2.8 \%} $ &   $ 10.215 (4 ) ^{+ 2.7 \%}_{ -2.2 \%} $ \\ [0.1cm]
nLOPS$_{\rm QCD}$  &  $ 0.924 (5 ) ^{+ 2.7 \%}_{ -2.4 \%}  $ &   $ 2.002 (2 ) ^{+ 3.2 \%}_{ -2.5 \%} $ &   $ 1.991 (1 ) ^{+ 3.1 \%}_{ -2.5 \%} $ &   $ 10.23(2) ^{+ 2.8 \%}_{ -2.3 \%} $ \\ [0.1cm]
NLOPS$_{\rm had}$  &  $ 0.8321 (3 ) ^{+ 3.0 \%}_{ -2.5 \%} $ &   $ 1.8110 (6 ) ^{+ 3.6 \%}_{ -2.9 \%} $ &   $ 1.8036 (6 ) ^{+ 3.5 \%}_{ -2.8 \%} $ &   $ 9.576 (3 ) ^{+ 2.7 \%}_{ -2.2 \%} $ \\ [0.1cm]
nLOPS$_{\rm had}$  &  $ 0.8481 (4 ) ^{+ 2.6 \%}_{ -2.4 \%} $ &   $ 1.8429 (8 ) ^{+ 3.1 \%}_{ -2.5 \%} $ &   $ 1.8374 (6 ) ^{+ 3.1 \%}_{ -2.5 \%} $ &   $ 9.460 (9 ) ^{+ 2.8 \%}_{ -2.2 \%} $ \\ [0.1cm]
nLO-MJM$_{\rm had}$            &  $ 0.963 (1 ) ^{+ 14.0 \%}_{ -6.7 \%} $ &   $ 2.093 (2 ) ^{+ 15.2 \%}_{ -7.3 \%} $ &   $ 2.074 (2 ) ^{+ 13.9 \%}_{ -7.0 \%} $ &   $ 10.32 (1 ) ^{+ 13.2 \%}_{ -6.4 \%} $ \\ [0.1cm]    
\hline
\end{tabular}
  \end{center}
  \caption{Fiducial cross sections in the setup described in Eq.~\ref{eq:fiducialvolume} for the doubly polarised processes.
  Uncertainties from 7-point QCD-scale variations are shown in percentages, while MC numerical uncertainties are shown in parentheses.
    \label{tab:matchedPOL}
  }
\end{table}
We stress that neither the LHE-level \pwhg~results more the truncated-shower \sher~ones are truly physical, but they have been introduced in this section for the sake of comparison with fixed order predictions (NLO QCD), in order to assess resummation effects in the absence of full QCD-shower effects. The physical results, \ie after showering, are presented in \refse{sec:ps_results}.

The broad validation at different perturbative orders presented in \refse{sec:fo_results} allows us to conclude that all DPA calculations agree very well at all considered orders, and the NWA and BW calculations from \sher~and \madg~deviate by less than 2\% from the DPA results, making them compatible within the intrinsic uncertainties of the approximations. 

\subsection{Matched and merged results}\label{sec:ps_results}

In this section, we present several matched and merged predictions in the fiducial ATLAS setup, considering as reference fixed-order calculations the NLO QCD and loop-induced results from \moca, NNLO QCD results from \stri, and NLO EW results from \bbmc. These predictions are combined additively and multiplicatively in different ways,  
\begin{eqnarray}
\rd\sigma^{\rm (+)}_{\rm NNLO} &=&    
\rd\sigma_{\rm LO}\,\left(1+\delta^{\rm NLO}_{\rm QCD}+\delta^{\rm NNLO}_{\rm QCD}+\delta^{\rm NLO}_{\rm EW}\right)\,,\nnb\\
\rd\sigma^{\rm (\times)}_{\rm NNLO} &=&    
\rd\sigma_{\rm LO}\,\left(1+\delta^{\rm NLO}_{\rm QCD}+\delta^{\rm NNLO}_{\rm QCD}\right)\left(1+\delta^{\rm NLO}_{\rm EW}\right)\,.
\end{eqnarray}

As for matched predictions, the \pwhg~\cite{Pelliccioli:2023zpd} and \sher~\cite{Hoppe:2023uux} results in the presence of pure-QCD effects, namely exact or approximate NLO QCD corrections matched to pure-QCD shower effects from {\sc Pythia 8} and {\sc Sherpa} showers at parton level, are denoted by NLOPS$_{\rm QCD}$ and nLOPS$_{\rm QCD}$ respectively. The cases of matching to a full QCD+QED shower, including hadronisation and MPIs, are labelled with the subscript ``had''. In addition to the PS matched predictions, we show predictions from a multi-jet merged sample obtained with ~\sher~(nLO-MJM$_{\rm had}$) \cite{Hoppe:2023uux}. In the multi-jet merged sample, the 0- and 1-jet samples are included at nLO QCD accuracy, while 2- and 3-jet contributions are included at LO. The loop-induced contributions are not included to separate the effect of higher-order corrections arising from hard radiation unless otherwise stated. 
The results for the loop-induced $\Pg\Pg$ contributions beyond fixed-order are discussed in \refse{sec:comp_exp}.

The upper part of Table \ref{tab:matchedPOL} shows the polarised fiducial cross sections for the combined QCD and EW calculations. We observe that the different combination schemes for NNLO QCD and NLO EW lead to the expected negative corrections with respect to the pure NNLO QCD result, with and without including the $\Pg\Pg$ contribution. We find corrections of $-7\%$ and $-10\%$ for the additive and multiplicative combination. The size and the difference between the combination schemes are largely polarisation independent.

Turning towards the matched computations and taking the fixed-order NLO QCD as a reference, we observe only minor corrections on the integrated level for \pwhg~shower of about two permille in the pure QCD shower setup. At the same time, we find differences of $+3\%$ for the nLOPS setup. These differences can be traced back to the differences in the on-shell approximation and are consistent with the expected internal uncertainty. The QCD+EW shower predictions receive a negative $-6\%$ correction, similar to the fixed-order EW corrections in the additive matching scheme.

The multi-jet merged predictions are substantially higher than the matched predictions and agree very well with the NNLO QCD results, confirming the picture that the source of higher-order QCD corrections  mainly arises from hard-radiation events. The scale uncertainties are multiple times larger than at NNLO QCD because of the merging procedure.

\begin{figure}
  \centering
  \subfigure[off-shell\label{fig:nlops_dphiee__ff}]{\includegraphics[width=0.49\textwidth, page=1]{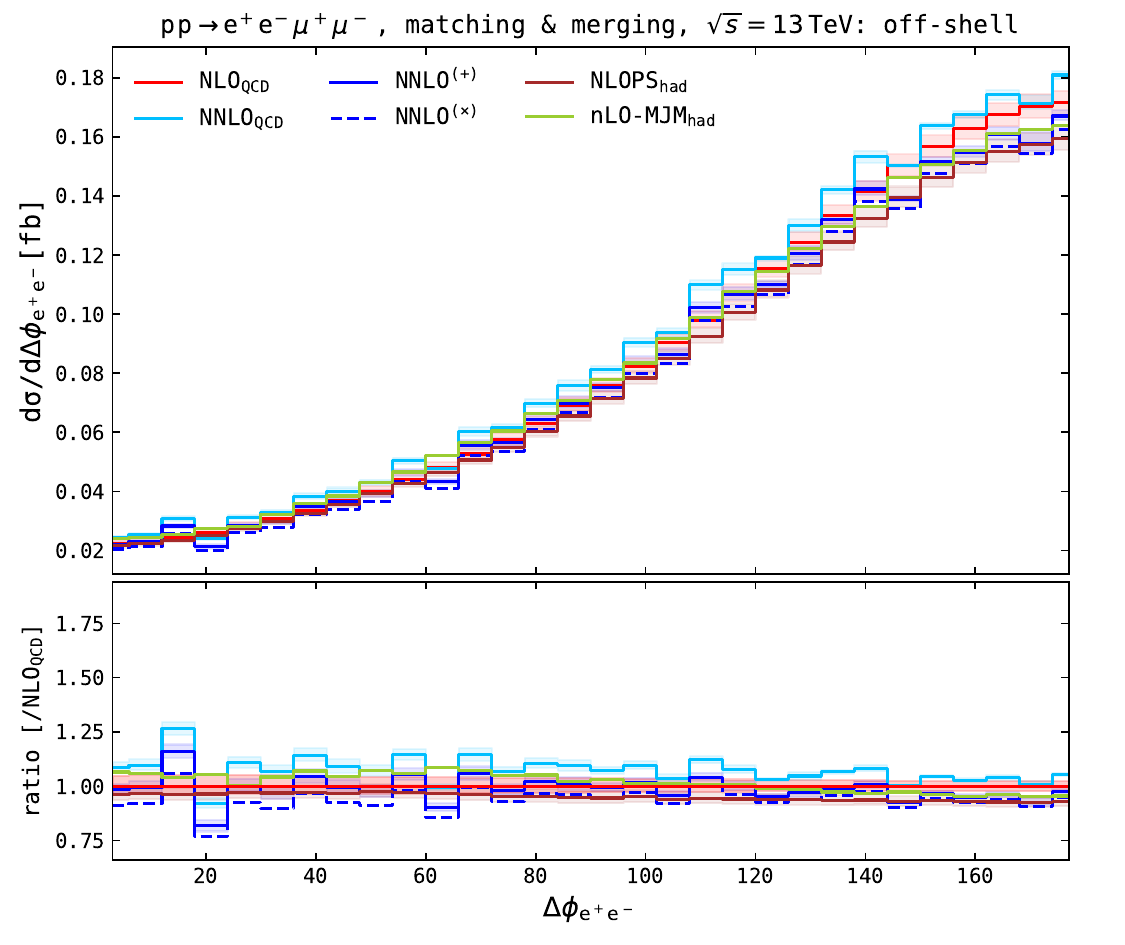}}
  \subfigure[unpol.   \label{fig:nlops_dphiee__uu}]{\includegraphics[width=0.49\textwidth, page=1]{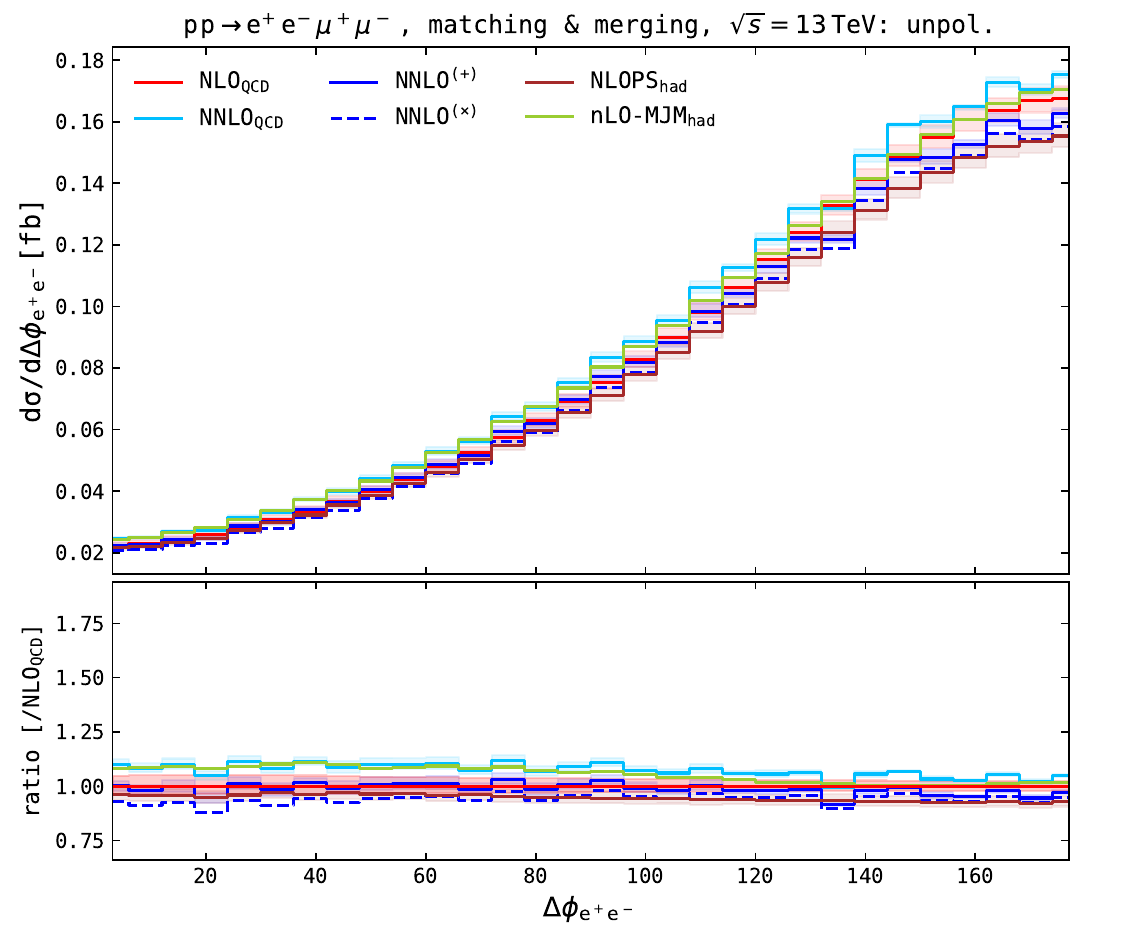}}\\
  \subfigure[LL       \label{fig:nlops_dphiee__ll}]{\includegraphics[width=0.49\textwidth, page=1]{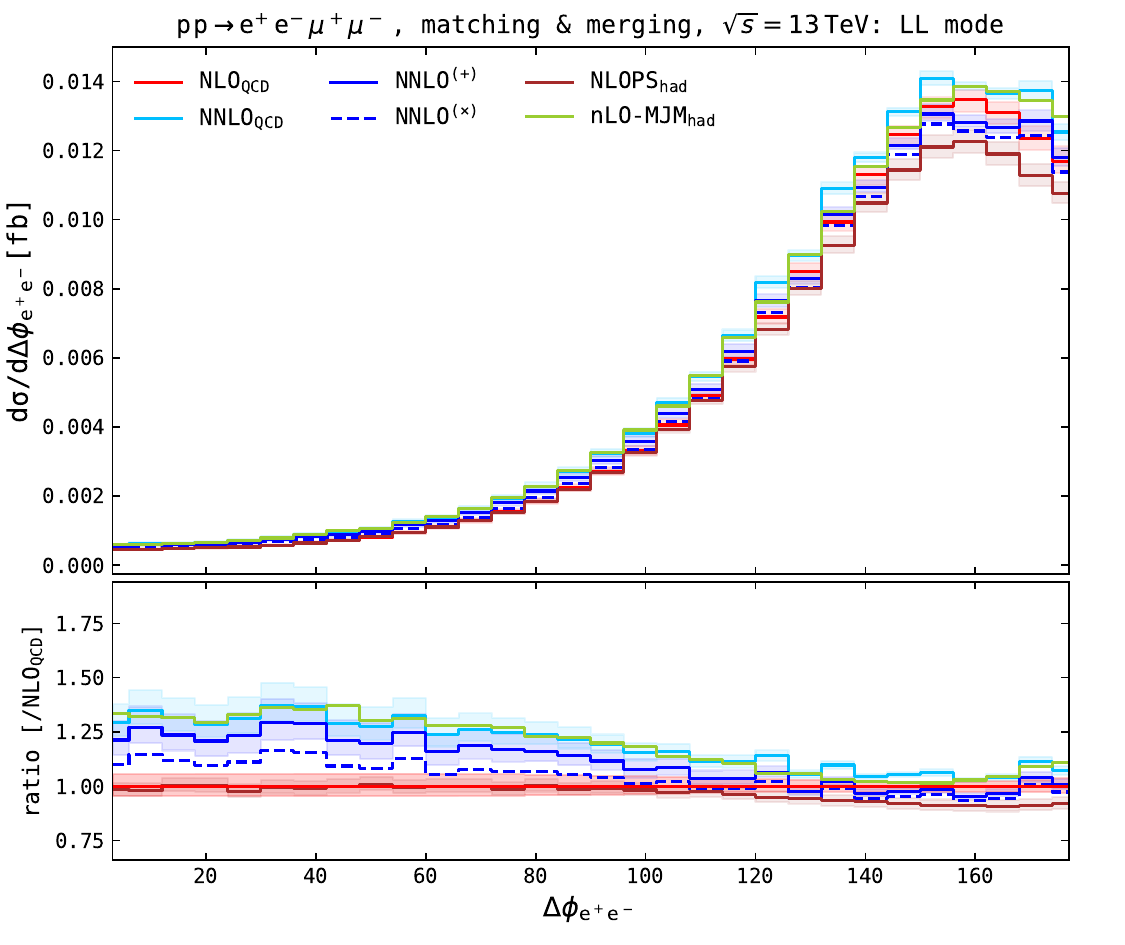}}
  \subfigure[LT       \label{fig:nlops_dphiee__lt}]{\includegraphics[width=0.49\textwidth, page=1]{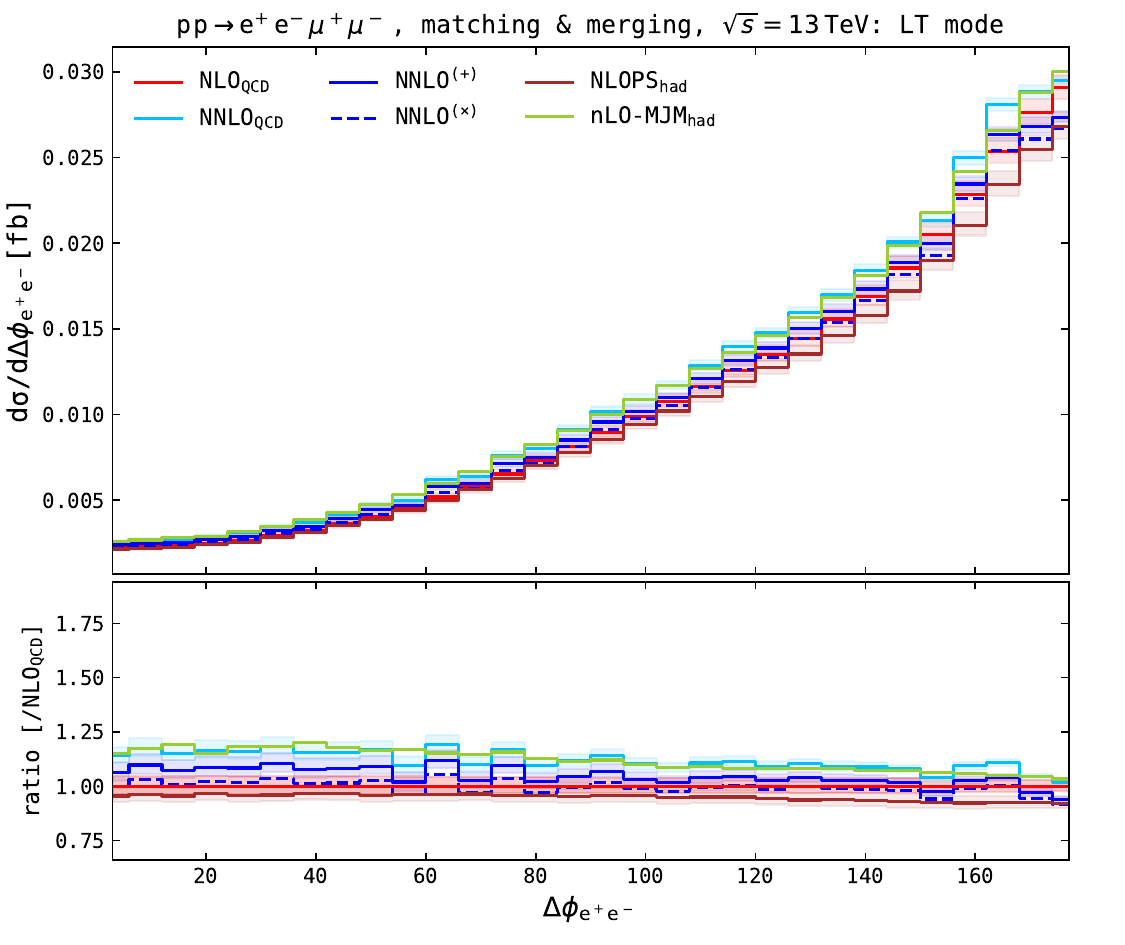}}\\
  \subfigure[TL       \label{fig:nlops_dphiee__tl}]{\includegraphics[width=0.49\textwidth, page=1]{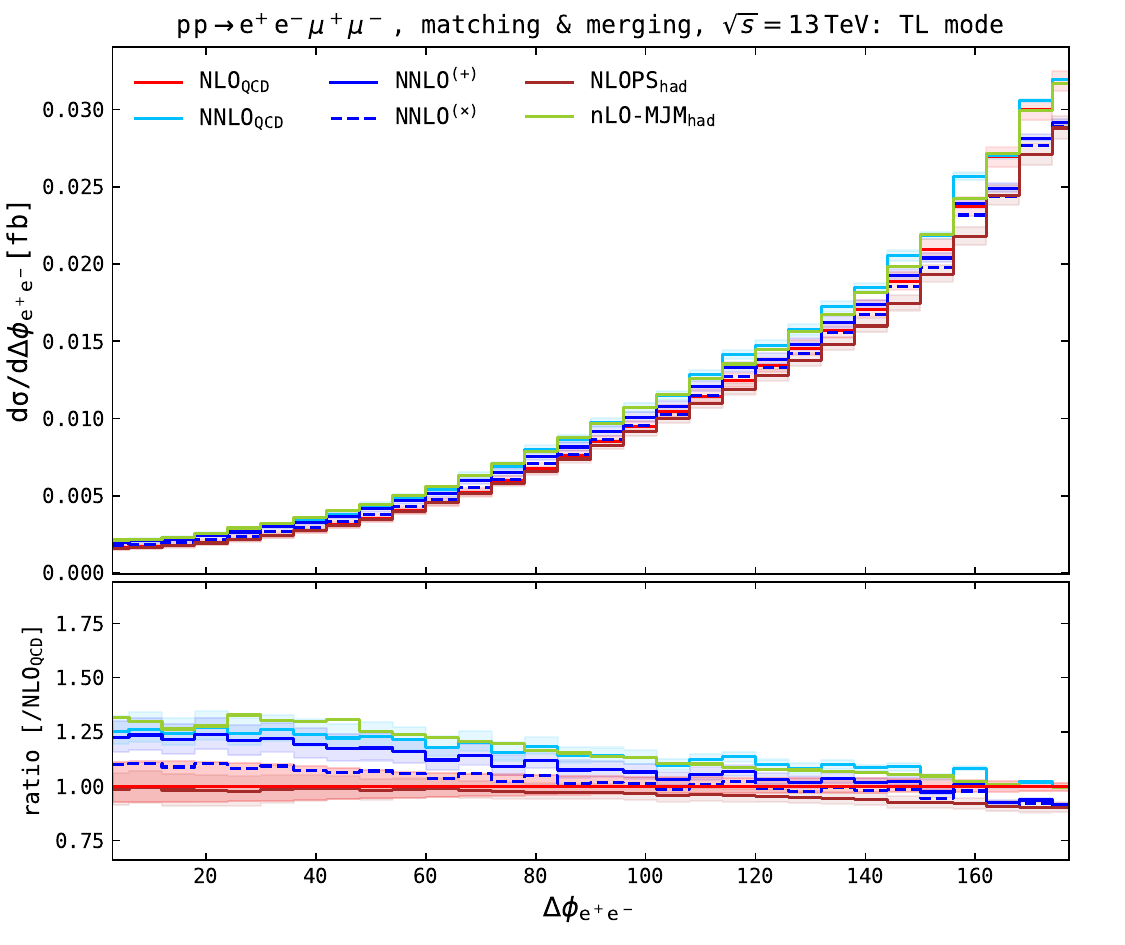}}
  \subfigure[TT       \label{fig:nlops_dphiee__tt}]{\includegraphics[width=0.49\textwidth, page=1]{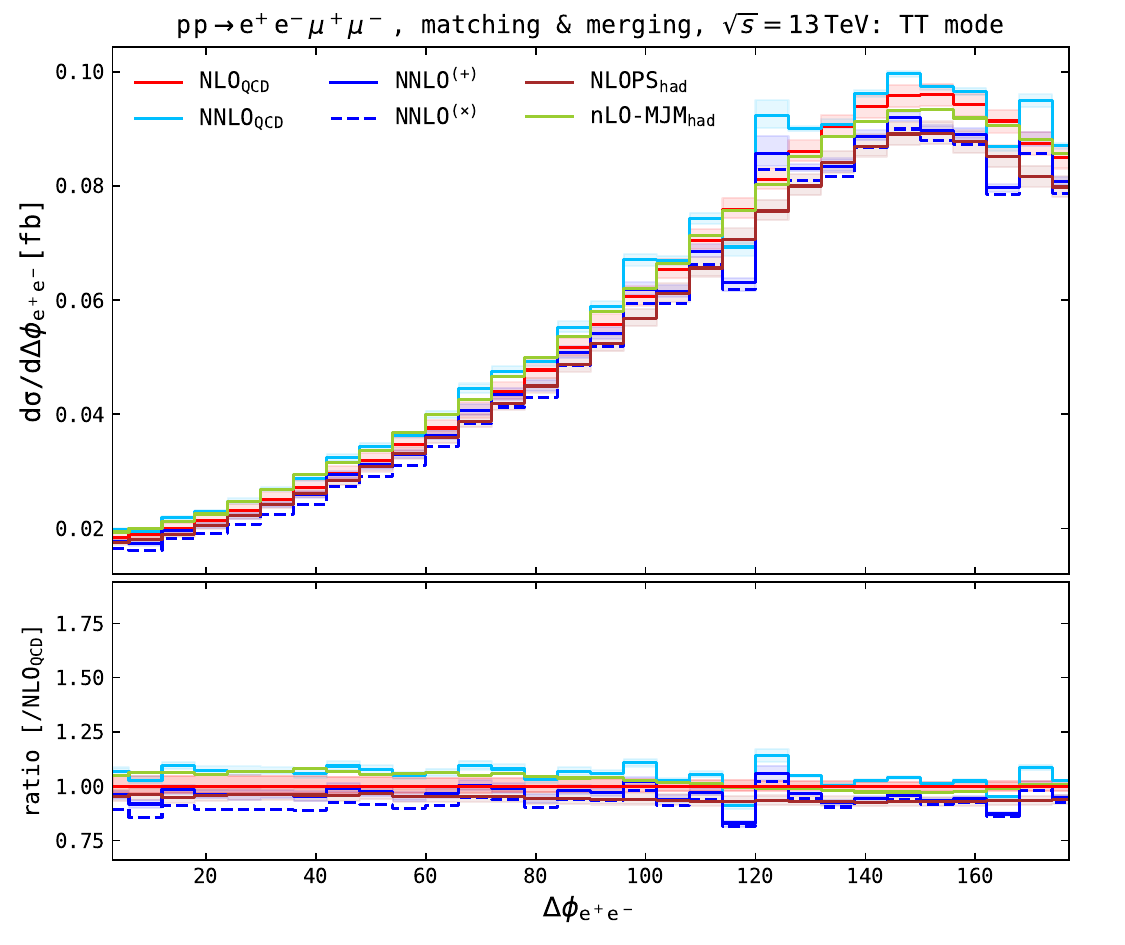}}
  \caption{
    Fiducial distributions in the positron-electron azimuthal distance.
    Absolute distributions are shown in main panels, ratios
    over fixed-order NLO QCD results are shown in lower panels.
    Shaded bands in the lower panels represent QCD-scale uncertainties.
  }\label{fig:NLOPS_dphiee}
\end{figure}

\begin{figure}
 \centering
   \subfigure[off-shell\label{fig:nlops_ptz__ff}]{\includegraphics[width=0.49\textwidth, page=2]{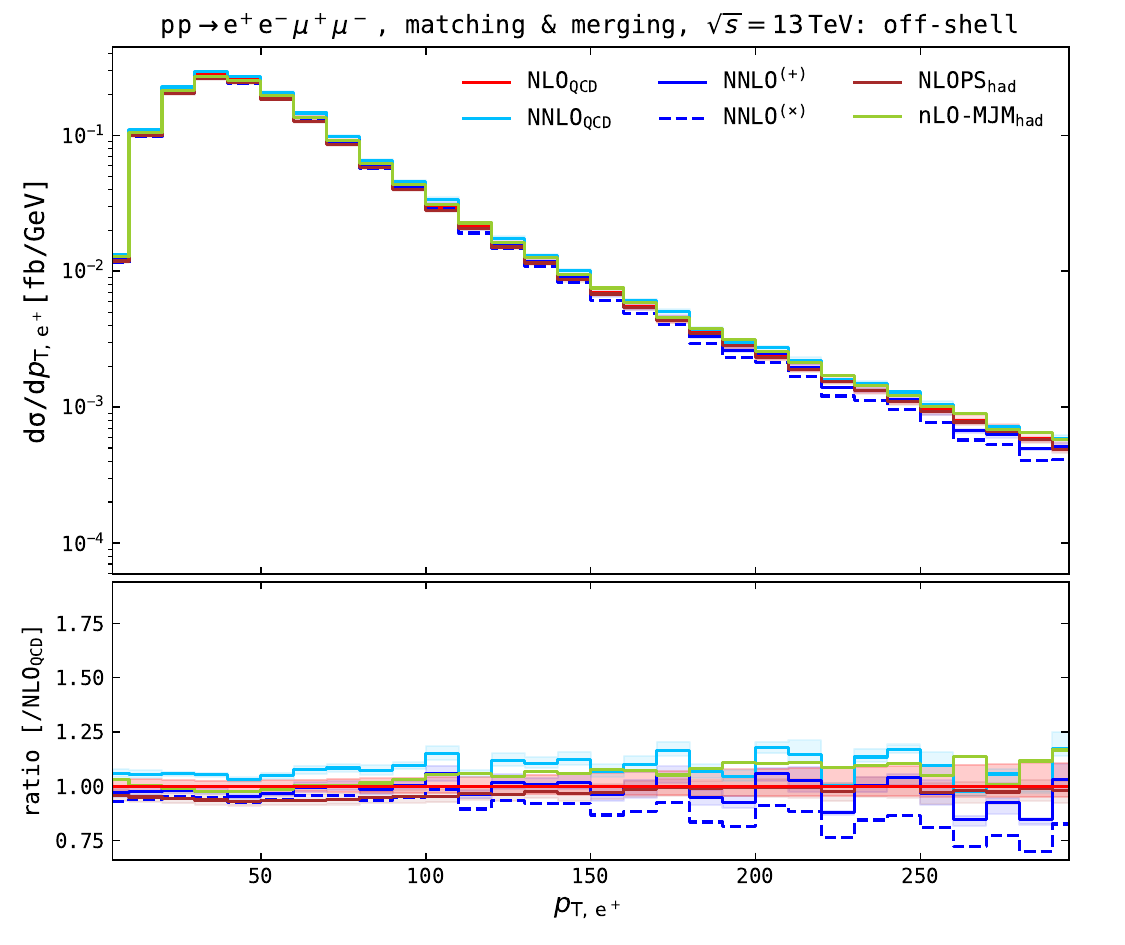}}
   \subfigure[unpol.   \label{fig:nlops_ptz__uu}]{\includegraphics[width=0.49\textwidth, page=2]{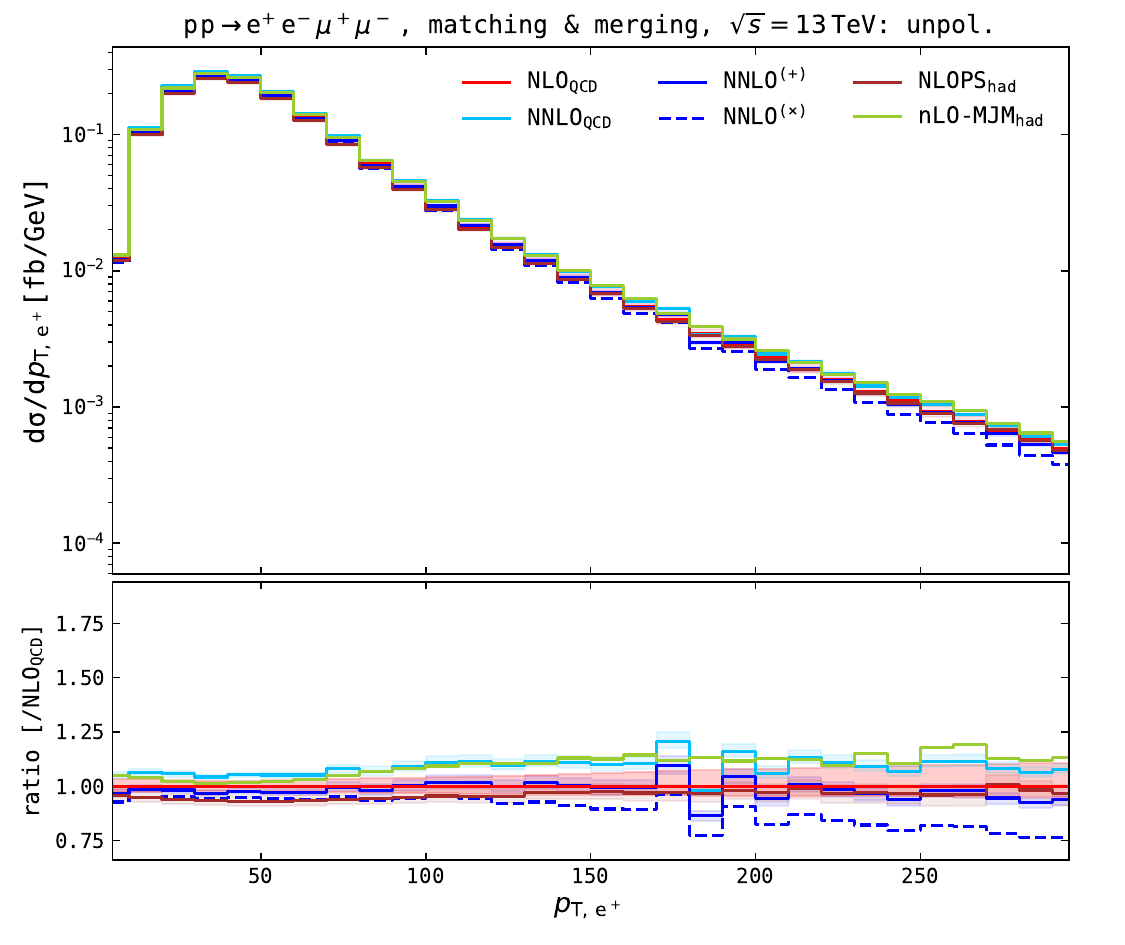}}\\
   \subfigure[LL       \label{fig:nlops_ptz__ll}]{\includegraphics[width=0.49\textwidth, page=2]{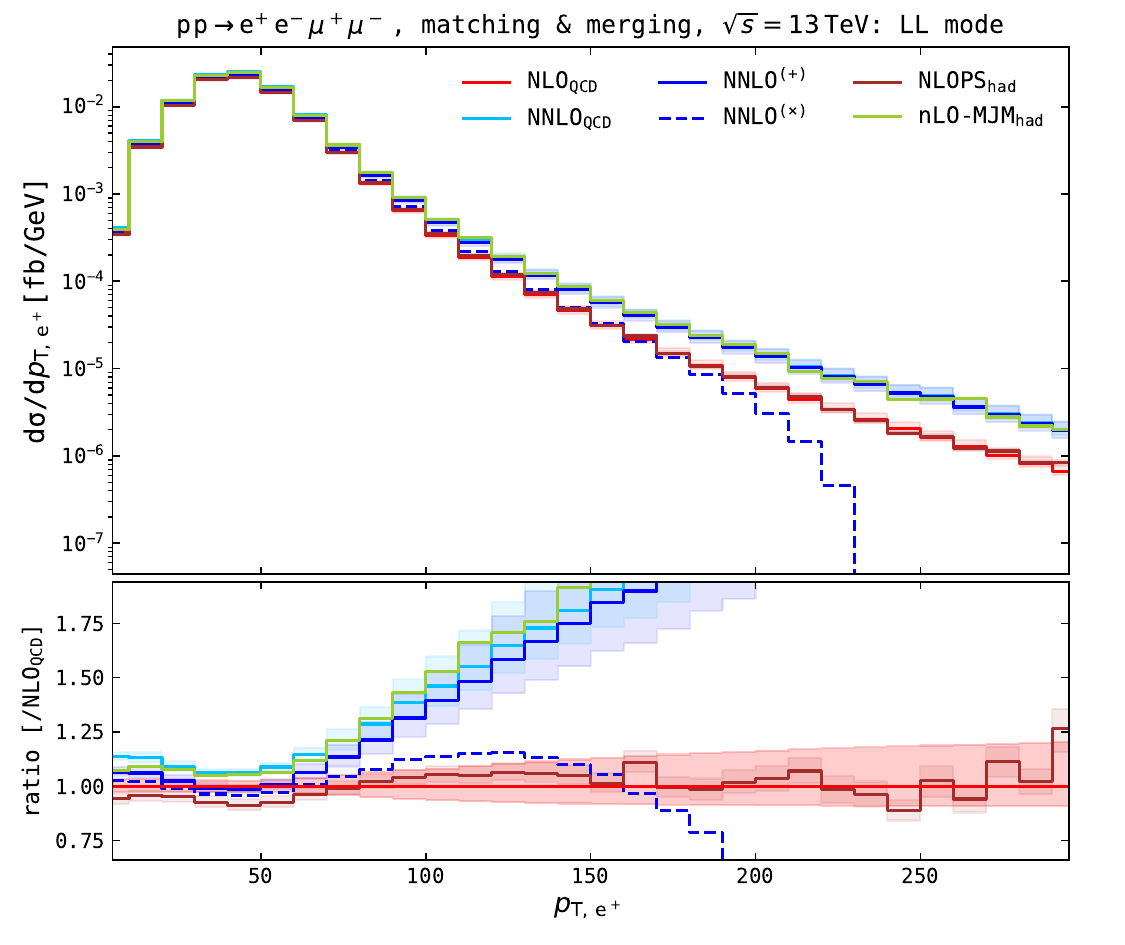}}
   \subfigure[LT       \label{fig:nlops_ptz__lt}]{\includegraphics[width=0.49\textwidth, page=2]{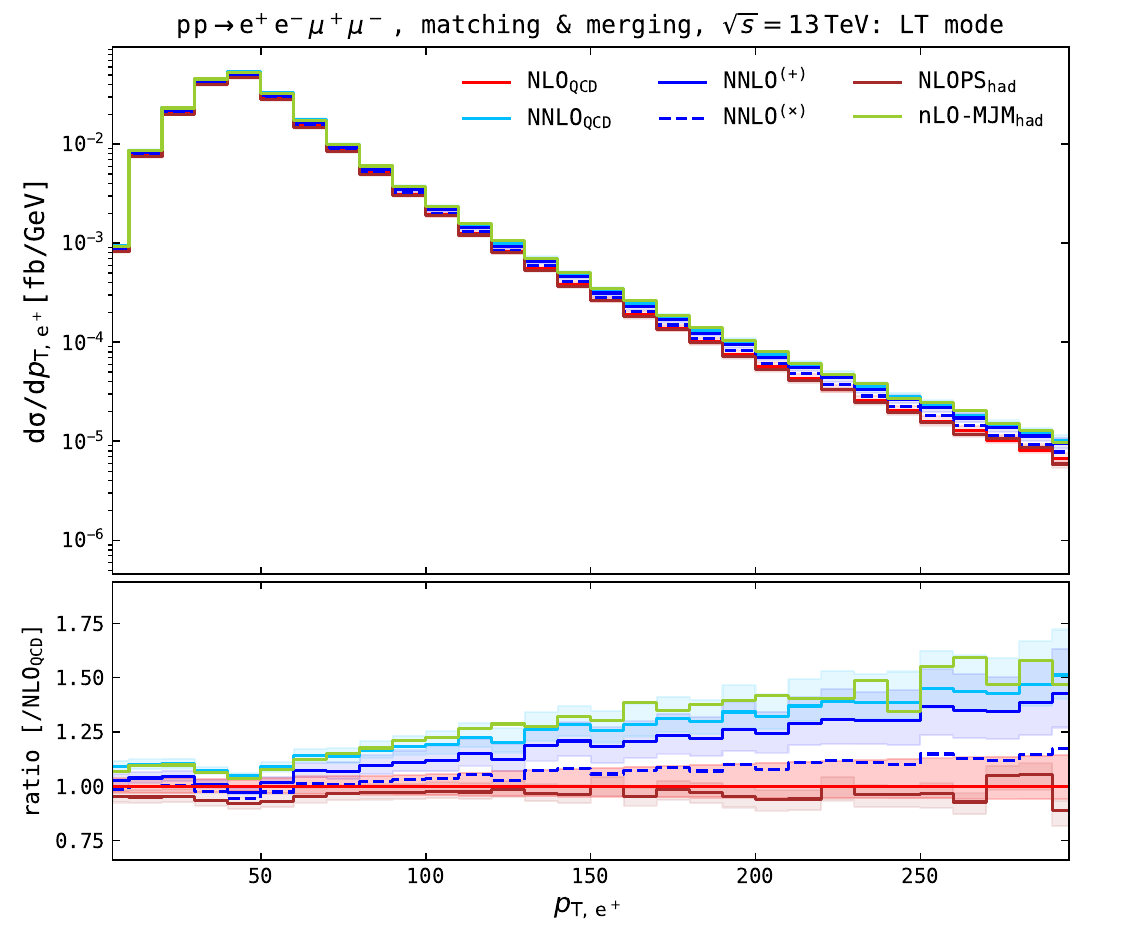}}\\
   \subfigure[TL       \label{fig:nlops_ptz__tl}]{\includegraphics[width=0.49\textwidth, page=2]{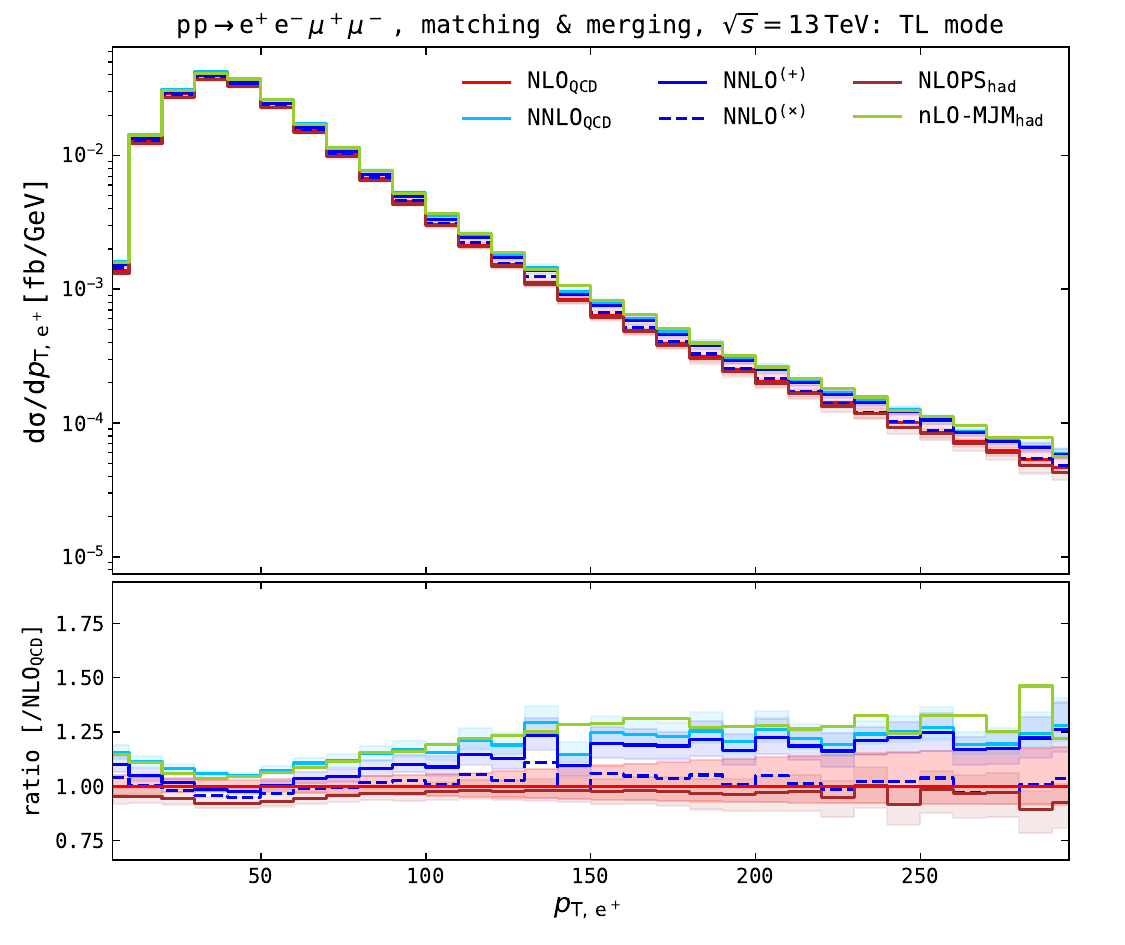}}
   \subfigure[TT       \label{fig:nlops_ptz__tt}]{\includegraphics[width=0.49\textwidth, page=2]{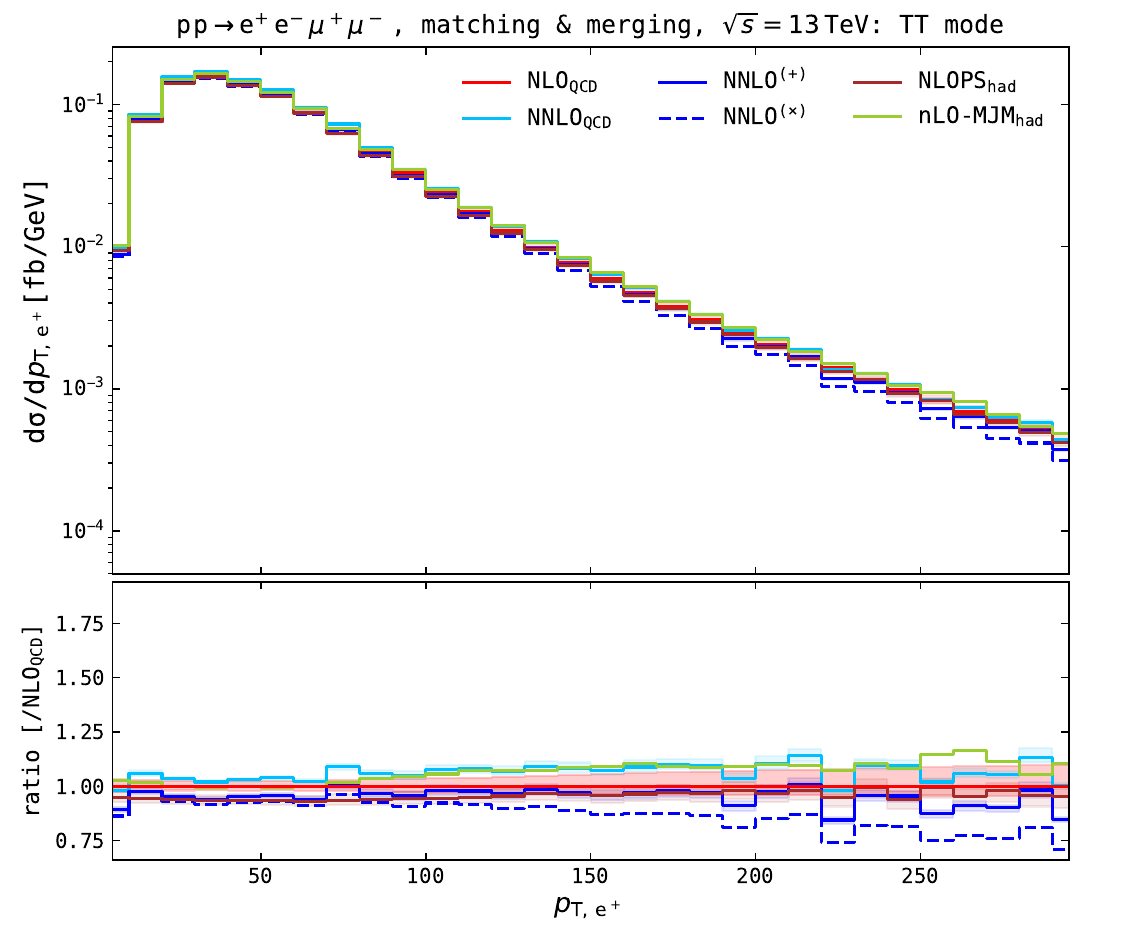}}
   \caption{
     Fiducial distributions in the positron transverse momentum.
     Same structure as Fig.~\ref{fig:NLOPS_dphiee}.
   }\label{fig:NLOPS_ptz}
 \end{figure}

On the differential level, we focus on the azimuthal opening angle between the electron and positron and the transverse momentum of the positron, shown in Fig.~\ref{fig:NLOPS_dphiee} and Fig.~\ref{fig:NLOPS_ptz}, respectively, for different polarisation setups. In both collections of plots, we present absolute results and the ratio to NLO QCD predictions to highlight the impact of the higher-order corrections.

For the NLOPS$_{\rm had}$ result, we observe overall negative parton-shower effects (including QED and non-perturbative contributions) with respect to NLO QCD dominantly at the bulk of the phase space, in agreement with the total cross section. It is known from fixed-order calculations that real QED corrections give a negative correction at the $\PZ$ pole, both in the off-shell description and in the DPA \cite{Denner:2021csi}. Similarly, the matching to QED-shower leads to a 6\% negative shift from the pure-QCD calculation \cite{Pelliccioli:2023zpd}, as also found for integrated cross sections (see Table~\ref{tab:matchedPOL}). The impact of hadronisation and MPI is much smaller, with sub-percent effects on differential distributions.

The pure NNLO QCD corrections depend on the polarisation setup. For the off-shell, the unpolarised and the doubly transverse polarised predictions, we observe corrections of $\order{7\%}$ similar to the total fiducial cross section, which are rather flat across the phase space. When at least one Z boson is longitudinally polarised, the corrections have visible shapes. For $\Delta\phi_{\Pe^+\Pe^-}$, the NNLO QCD corrections increase towards smaller angles, while for the transverse momentum, we see enhancements for large $\pt{\Pe^+\Pe^-}$. Both observations can be explained through the extra hard emissions included in these calculations since a small angle corresponds to boosted, i.e. high $p_{\rT}$ Z bosons. The corrections reach up to $\sim25\%$ for for small $\Delta\phi_{\Pe^+\Pe^-}$ and $\sim100\%$ for $\pt{\Pe^+\Pe^-} \approx 250$ GeV.

The NNLO QCD+NLO EW predictions show similar behaviour except that the typical electroweak Sudakov logarithms appear in the $\pt{\Pe^+\Pe^-}$ distribution. A small imprint of this negative correction at larger $p_{\rT}$ can also be seen in the $\Delta\phi_{\Pe^+\Pe^-}$ distribution. These effects counteract the positive corrections of the NNLO QCD corrections in cases with longitudinal Z bosons. The effect is more significant in the multiplicative combination approach. In the double longitudinally polarised case, the multiplicative matched approach renders the cross section negative at high $\pt{\Pe^+\Pe^-}$, aligning with the same behaviour at NLO EW. This is an artefact of the small LO cross-section, which renders $\delta^{\rm NLO}_{\rm EW}$ smaller than $-1$. In the additive case, the significant and positive higher-order QCD corrections compensate for this.

Finally, we turn to the multi-jet merged predictions from \sher. The first observation we can make is that the central multi-jet merged prediction aligns very well with the central NNLO QCD prediction, in particular in phase space regions where the NNLO QCD corrections are significant, which further supports the interpretation that the large NNLO QCD corrections arise from hard radiation. This observation holds for the unpolarised and doubly polarised cases, in the off-shell case, the normalisation is about $5\%$ smaller for the \sher~prediction.

In summary, we find good agreement between NLO QCD and NLO+PS predictions, indicating that the overall effect of parton showers on the differential cross sections is small. The differences can largely be attributed to the QED effects included in the shower. More significant differences arise when additional contributions from hard radiation are included, as in the case of NNLO QCD or multi-jet merged predictions. The electroweak corrections are negative and can be separated into two regions: first, the bulk of the phase space receives a $-6\%$ correction, and, second, the typical Sudakov logarithms arise in the high-energy limit.

\subsection{Joint polarisation fractions}\label{sec:fractions}
In this section we study how the theoretical modelling affects the joint polarisation fractions which are extracted experimentally. Such fractions are defined as ratios of doubly polarised cross sections over the unpolarised one,
\beq\label{eq:frac_def}
f_{\lambda\lambda'} = \frac{\sigma_{\lambda\lambda'}}{\sigma_{\rm unp}}\,,\qquad \lambda,\lambda' = \rL,\rT\,.
\eeq
Note that the sum of polarisation fractions differs from 1 by interference contributions, which do not vanish owing to selection cuts but are small in typical LHC fiducial volumes. 
In the ATLAS measurement \cite{ATLAS:2023zrv}, the off-shell effects are considered as a background and therefore not considered as a separate template in the polarised-template fit.
We note here that the different definitions of the polarised cross sections using the DPA, NWA, or BW are associated with different off-shell contributions.
As the off-shell contribution is subtracted as a background in the ATLAS measurement \cite{ATLAS:2023zrv}, it would be difficult to interpret the measured polarisation fractions if the off-shell contribution is not precisely given.
At the present level of accuracy, this issue is irrelevant as the off-shell effects are very small compared to the other uncertainties.
However, for future precise measurements of polarisation fractions, it may be better to treat off-shell effects as a separate template and provide this information together with the polarisation templates (including a template for the polarisation interference) to avoid misleading
interpretation of the polarisations.

In Table~\ref{tab:fractions} we show joint polarisation fractions obtained at different fixed perturbative orders as well as including PS effects in the ATLAS fiducial setup.
In the case of perturbative modelling, we have introduced the combination of results in the $q\overline{q}$ and $\Pg\Pg$ channels as follows:
\begin{eqnarray}
\rd\sigma^{\rm (+)}_{\rm NNLO,\,gg} &=& \rd\sigma^{\rm (+)}_{\rm NNLO} + \rd\sigma_{\Pg\Pg} \,,  \nnb\\ 
\rd\sigma^{\rm (\times)}_{\rm NNLO,\,gg} &=& \rd\sigma^{\rm (\times)}_{\rm NNLO} + \rd\sigma_{\Pg\Pg} \,.
\end{eqnarray}

\begin{table}
  \begin{center}
    \begin{tabular}{lccccc}
      \hline\rule{0ex}{2.7ex}
\cellcolor{blue!9} & \cellcolor{blue!9}LL [\%] & \cellcolor{blue!9}LT [\%]& \cellcolor{blue!9}TL [\%]& \cellcolor{blue!9}TT [\%] & \cellcolor{blue!9}interf [\%]\\
\hline\\[-0.25cm]
NLO$_{\rm QCD }$  &   $ 5.87 ^{+ 0.03   }_{ -0.05   }$ &   $ 12.74 ^{+ 0.07   }_{ -0.06   }$ &   $ 12.69 ^{+ 0.07   }_{ -0.06   }$ &   $ 67.35 ^{+ 0.14   }_{ -0.16   }$ &   $ 1.35 $ \\[0.1cm]
NNLO$_{\rm QCD }$  &   $ 6.07 ^{+ 0.05   }_{ -0.04   }$ &   $ 13.11 ^{+ 0.09   }_{ -0.08   }$ &   $ 13.04 ^{+ 0.08   }_{ -0.07   }$ &   $ 66.20 ^{+ 0.19   }_{ -0.24   }$ &   $ 1.58 $ \\[0.1cm]
NNLO$^{\rm (+)}$  &   $ 6.12 ^{+ 0.06   }_{ -0.04   }$ &   $ 13.29 ^{+ 0.09   }_{ -0.08   }$ &   $ 13.21 ^{+ 0.08   }_{ -0.07   }$ &   $ 65.75 ^{+ 0.20   }_{ -0.25   }$ & 1.63\\[0.1cm]
NNLO$^{\rm (\times)}$  &   $ 6.12 ^{+ 0.05   }_{ -0.04   }$ &   $ 13.24 ^{+ 0.09   }_{ -0.08   }$ &   $ 13.16 ^{+ 0.08   }_{ -0.07   }$ &   $ 65.92 ^{+ 0.19   }_{ -0.24   }$ & 1.56 \\[0.1cm]
NNLO$^{\rm (+)}_{\rm gg}$  &   $ 6.05 ^{+ 0.03   }_{ -0.03   }$ &   $ 12.15 ^{+ 0.10   }_{ -0.15   }$ &   $ 12.07 ^{+ 0.11   }_{ -0.16   }$ &   $ 68.29 ^{+ 0.30   }_{ -0.21   }$ & 1.44 \\[0.1cm]
NNLO$^{\rm (\times)}_{\rm gg}$  &   $ 6.04 ^{+ 0.03   }_{ -0.02   }$ &   $ 12.06 ^{+ 0.11   }_{ -0.16   }$ &   $ 11.99 ^{+ 0.12   }_{ -0.17   }$ &   $ 68.53 ^{+ 0.33   }_{ -0.23   }$ & 1.38\\[0.1cm] 
\hline\\[-0.25cm]
NLOPS$_{\rm QCD }$  &   $ 5.88 ^{+ 0.03   }_{ -0.04   }$ &   $ 12.76 ^{+ 0.08   }_{ -0.06   }$ &   $ 12.71 ^{+ 0.07   }_{ -0.06   }$ &   $ 67.30 ^{+ 0.13   }_{ -0.15   }$ &   $ 1.35 $ \\[0.1cm]
NLOPS$_{\rm had }$  &   $ 5.86 ^{+ 0.03   }_{ -0.04   }$ &   $ 12.74 ^{+ 0.08   }_{ -0.06   }$ &   $ 12.69 ^{+ 0.07   }_{ -0.06   }$ &   $ 67.38 ^{+ 0.13   }_{ -0.15   }$ &   $ 1.33 $ \\[0.1cm]
nLOPS$_{\rm QCD }$  &   $ 6.02 ^{+ 0.05   }_{ -0.08   }$ &   $ 13.04 ^{+ 0.04   }_{ -0.09   }$ &   $ 12.97 ^{+ 0.04   }_{ -0.09   }$ &   $ 66.61 ^{+ 0.14   }_{ -0.47   }$ &   $ 1.36 $ \\[0.1cm]
nLOPS$_{\rm had }$  &   $ 5.98 ^{+ 0.03   }_{ -0.07   }$ &   $ 12.99 ^{+ 0.02   }_{ -0.09   }$ &   $ 12.96 ^{+ 0.02   }_{ -0.09   }$ &   $ 66.70 ^{+ 0.22   }_{ -0.46   }$ &   $ 1.37 $ \\[0.1cm]
nLO-MJM$_{\rm had}$  &   $ 6.14 ^{+ 0.12   }_{ -0.11   }$ &   $ 13.35 ^{+ 0.47   }_{ -0.32   }$ &   $ 13.23 ^{+ 0.32   }_{ -0.28   }$ &   $ 65.85 ^{+ 1.11   }_{ -0.96   }$ &   $ 1.43 $ \\[0.1cm]
\hline\\[-0.25cm]
pre-fit \cite{ATLAS:2023zrv} &   $ 6.1\pm 0.4 $ &   \multicolumn{2}{c}{$ 22.9\pm0.9$} &   $ 69.9\pm3.9 $ & 1.1\\[0.1cm]
post-fit \cite{ATLAS:2023zrv} &   $ 7.1 \pm 1.7 $ &   \multicolumn{2}{c}{$22.8\pm1.1 $}  &   $ 69.0\pm2.7$ & 1.1\\[0.1cm]
\hline
    \end{tabular}\qquad
  \end{center}
  \caption{Fiducial polarisation fractions (in percentages) in the setup described in Eq.~\ref{eq:fiducialvolume}, obtained with various perturbative orders and with different matching or merging procedures. Uncertainties are obtained from correlated 7-point QCD-scale variations. The last two rows represent the estimated SM value (pre-fit) and the fit results (post-fit) of the ATLAS analysis \cite{ATLAS:2023zrv}, respectively. The LT and TL fractions are summed in these last rows.
    \label{tab:fractions}   
  }
\end{table}
The interference contribution amounts at about 1.4\%, with small variations owing to different orders and calculation details. 
The ATLAS pre-fit results \cite{ATLAS:2023zrv} were obtained using \moca~predictions \cite{Denner:2021csi} which include NLO QCD and NLO EW corrections combined with the LO prediction for the $\Pg\Pg$ channel.
The dependence of polarisation fractions on higher-order corrections is not as marked as for integrated and differential polarised cross sections. In particular, we observe that the inclusion of NNLO QCD corrections enhances by about 4\% the LL and mixed fractions, consequently reducing the TT one. The NLO EW corrections give another 1\% enhancement to the LL and mixed polarisation fractions. Conversely, the gluon-induced contribution is dominated by the TT state (90\%), leading to a 4\% enhancement of the TT fraction.
For our best fixed-order prediction, obtained by combining additively the NNLO QCD and NLO EW corrections and including the loop-induced $\Pg\Pg$ contribution, we also computed the uncertainties from 31-point QCD-scale variations\footnote{The renormalisation and factorisation scales are independently varied by factors of 2 and 1/2 in the numerator and in the denominator of Eq.~\ref{eq:frac_def}, excluding the cases where scale-factor ratios between numerator and denominator are either 4 or 1/4.}. With this approach, the QCD uncertainties associated with joint fractions are more conservative than those found with correlated scale variations: 
\begin{eqnarray}
f_{\rL\rL} &= 6.05 ^{+ 0.03  \,(+0.27  )   }_{ -0.03  \,(-0.24   )   }\%\,,\quad
f_{\rL\rT} &= 12.15 ^{+ 0.10  \,(+0.39  )   }_{ -0.15 \, (-0.24   )   }\%\,,\nnb\\
f_{\rT\rL} &= 12.07 ^{+ 0.11  \,(+0.39  )   }_{ -0.16 \, (-0.24   )   }\%\,,\quad
f_{\rT\rT} &= 68.29 ^{+ 0.30  \,(+3.04  )   }_{ -0.21 \, (-0.24   )   }\%\,,    
\end{eqnarray}

where the first uncertainties are correlated (i.e. we use the same scale choices in the numerator and denominator) while the second ones are obtained with 31-point scale variations. The ATLAS post-fit results agree with such predictions within the stated uncertainties. 
The matching to PS effects at exact or approximate NLO QCD accuracy (NLOPS from \pwhg, nLOPS from \sher) has a smaller impact than genuine NNLO QCD corrections, with at most 2\% modifications of the joint polarisation fractions.
The NLO merging from \sher~(nLO-MJM${}_{\rm had}$) approximates very well the combined NNLO$^{(\times/+)}$ results for joint fractions, showing again the importance of including NLO QCD corrections to the $\PZ\PZ\Pj$ process in event generators.

\subsection{Comparison to ATLAS simulations}
\label{sec:comp_exp}
As already partially mentioned, the LHC Run-2 paradigm for the experimental extraction of either polarisation fractions or polarised cross sections \cite{Aaboud:2019gxl,Sirunyan:2020gvn,CMS:2021icx,ATLAS:2022oge,ATLAS:2023zrv,ATLAS:2024qbd,ATLAS:2025wuw} is represented by a data fit with separate SM templates for each (single or double) polarisation state of the di-boson signal, possibly accounting also for interference and non-resonant effects as additional templates or backgrounds.

\begin{figure}[ht]
   \centering
   \subfigure[$\cos\theta_{\Pe^+}^*$\label{fig:best_costhetastar_qq}]{\includegraphics[width=0.49\textwidth, page=4]{fig/our_best_predictions_qq.pdf}}
   \subfigure[$|y_{\Pe^+}-y_{\mu^+\mu^-}|$\label{fig:dyez_qq}]{\includegraphics[width=0.49\textwidth, page=1]{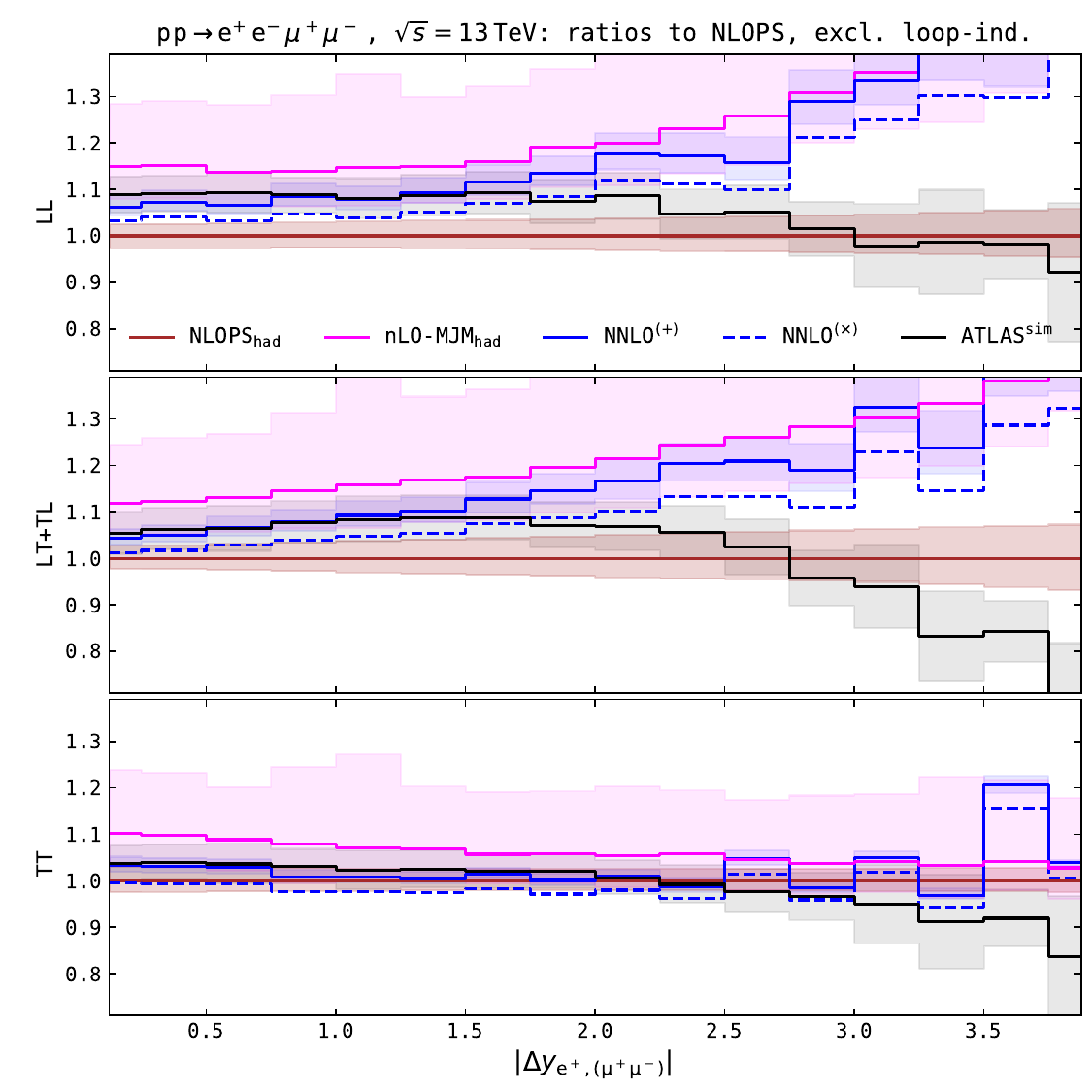}}
   \caption{
     Differential ratios over NLOPS$_{\rm had}$ results for the LL state (top panels), the sum of mixed states (middle panels) and the TT state (bottom panels), in the fiducial ATLAS setup described in Eq.~\ref{eq:fiducialvolume}. 
     The loop-induced $\Pg\Pg$ contribution is excluded from all predictions.
     The cosine of the positron decay angle (a)
     and the absolute value of the rapidity separation between the positron and the muon-antimuon system (b)
     are considered. Shaded bands from theoretical predictions come from QCD-scale variations, while those of the ATLAS predictions come from the envelope of all considered theoretical systematic uncertainties.
   }\label{fig:best_results}
 \end{figure}

In this section we compare the best predictions we have introduced in the previous sections to the SM predictions simulated by ATLAS for the Run-2 measurement \cite{ATLAS:2023zrv,ATL-PHYS-PUB-2025-021}. It is essential to recall that at the time of the analysis event generators capable of simulating polarised bosons as intermediate states were only available at LO accuracy, therefore higher-order perturbative effects have been included as correction factors to LO SM simulations by means of tailored reweighting techniques. In the following we consider ATLAS SM predictions for the $q\bar{q}$- and $\Pg\Pg$-induced production of polarised $\PZ$-boson pairs, while the EW production in association to two jets, \ie vector-boson scattering, is excluded.
The following theoretical systematic uncertainties affect the ATLAS predictions for doubly polarised signals \cite{ATLAS:2023zrv,ATL-PHYS-PUB-2025-021}:
\begin{itemize}
    \item[-] variations of PDF sets and of the strong coupling constant $\alphas$,
    \item[-] QCD-scale dependence, evaluated via 7-point scale variations around the central value for the factorisation and renormalisation scales (equal to the $\PZ$-boson pole mass),
    \item[-] inclusion of higher-order QCD corrections via reweighting of the LO-merged \madg~simulations for $q\bar{q}\rightarrow \PZ\PZ$, according to polarised \moca~K-factors for NLO QCD corrections and to NLO-merged \sher~unpolarised predictions for PS effects,
    \item[-] inclusion of NLO EW effects via reweighting LO-merged \madg~predictions with EW corrections obtained with \moca~(uncertainty as the difference between multiplicative and additive combination of NLO QCD and EW corrections),
    \item[-] modelling of interference effects by reweighting unpolarised NLO-merged \sher~predictions according to the interference-term \moca~prediction,
    \item[-] non-closure of the one-dimensional reweighting due to higher-order corrections,
    \item[-] inclusion of higher-order QCD effects to $\Pg\Pg\rightarrow \PZ\PZ$ through a reweighting of \sher~unpolarised predictions (LO-merged rescaled with flat NLO QCD corrections) according to LO \moca~polarised predictions.
\end{itemize}
The numerical results for polarised simulations used for the polarised-template fit of \citere{ATLAS:2023zrv} have been released in a recent publication note \cite{ATL-PHYS-PUB-2025-021}.
In \reffi{fig:best_results} these results are compared to a number of predictions that have been presented in the previous sections of this work. For clarity, we consider the loop-induced contribution separately.
On top of ATLAS-simulation results, the additive and multiplicative combinations of NNLO QCD (\stri) and NLO EW (\moca) corrections, and the approximate NLO-merged results (\sher) are shown as ratios to the NLOPS$_{\rm had}$ results (\pwhg). For angular decay observables the ATLAS modelling follows quite well the NLOPS results at the level of normalised shapes, as suggested by the rather flat ratios (black curves) in \ref{fig:best_costhetastar_qq}. Slight deviations only show up for the LL signal at the distribution end-points, where the LL cross section is expected to vanish. In terms of the normalisation, the ATLAS simulations show a fair agreement with combined NNLO QCD $+/\times$ NLO EW predictions, lying in between the multiplicative and additive combination in most of the distribution range. This shows the goodness of the reweighting procedure of LO-merged polarised predictions according to NLO QCD $+/\times$ NLO EW \moca~predictions \cite{Denner:2021csi}. 
For other observables, like the rapidity separation between the positron and the muon--antimuon pair shown in \reffi{fig:dyez_qq}, the ATLAS modelling is poorer, mostly due to the absence of polarisation effects for higher jet multiplicities that cannot be captured by NLO QCD corrections. This especially holds for the LL and mixed signals, which receive the largest corrections from hard QCD radiation. In this case, the modelling would benefit from NLO merging (even though approximate in virtual QCD effects), as shown by the shape-wise agreement between the nLO-MJM$_{\rm had}$ and combined NNLO predictions. The lack of virtual QCD corrections makes the NLO-merged results deviate from the NNLO ones by 5-to-10\% in most populated phase-space regions, and to much wider QCD-scale uncertainties (at the 15\% level, to be compared to 2\% in the NNLO predictions). The level of agreement between approximate NLO-merged and NNLO results somewhat improves for large transverse momenta and large rapidity separations, where the NNLO QCD calculation is dominated by hard real radiation (especially for the LL mode).

\begin{figure}[ht]
   \centering
   \subfigure[$\cos\theta_{\Pe^+}^*$\label{fig:best_costhetastar_gg}]{\includegraphics[width=0.49\textwidth, page=4]{fig/atlas_vs_th_gg.pdf}}
   \subfigure[$|y_{\Pe^+}-y_{\mu^+\mu^-}|$\label{fig:dyez_gg}]{\includegraphics[width=0.49\textwidth, page=1]{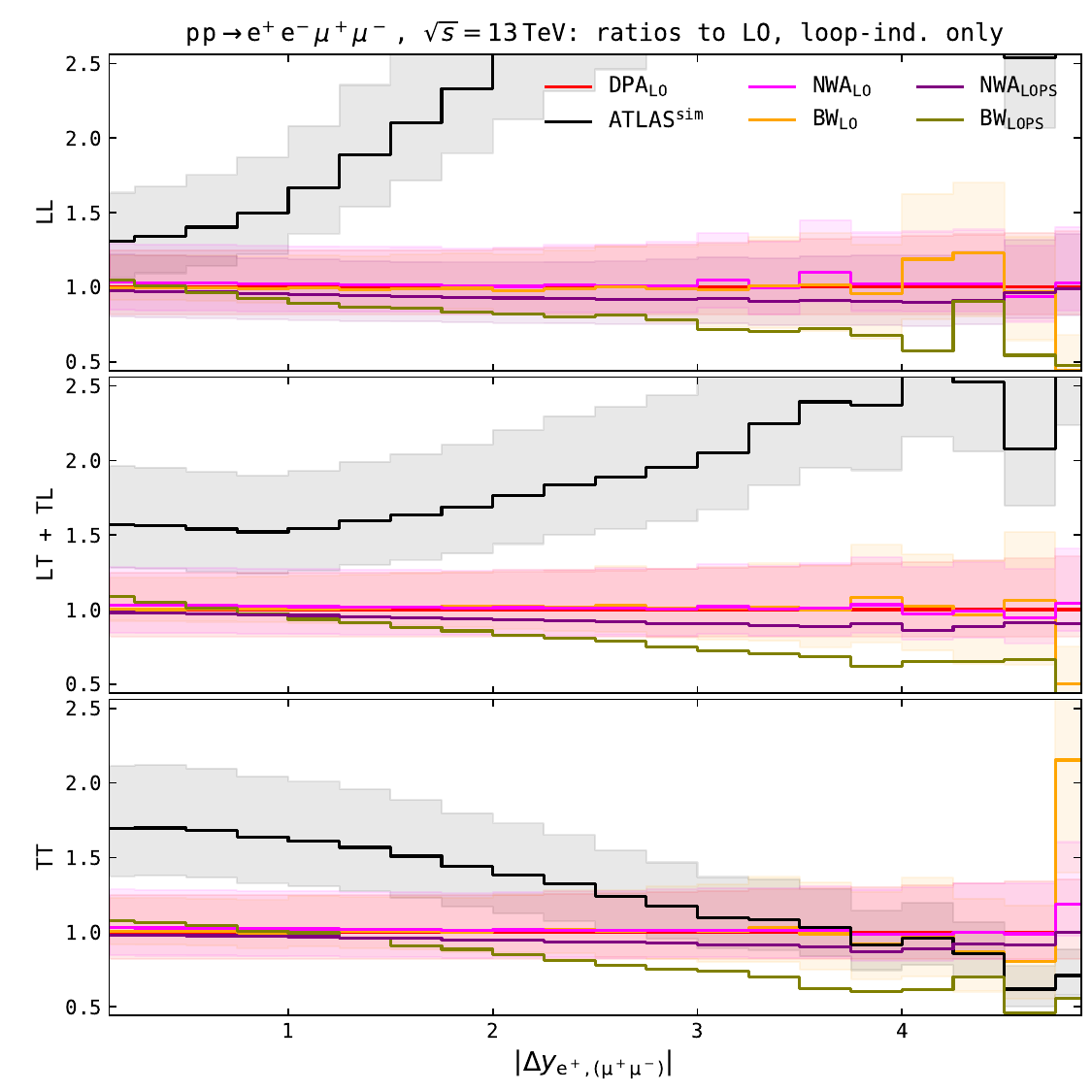}}
   \caption{
     Differential ratios over DPA$_{\rm LO}$ results for the loop-induced $\Pg\Pg$ contribution to the LL state (top panels), the sum of mixed states (middle panels) and the TT state (bottom panels), in the fiducial ATLAS setup described in Eq.~\ref{eq:fiducialvolume}. 
   Same structure and observables as in \reffi{fig:best_results}.
   }\label{fig:best_results_gg}
 \end{figure}

The $\Pg\Pg$ contribution to polarised $\PZ\PZ$ production, formally of the perturbative order accuracy as the genuine QCD corrections to the $q\bar{q}$ channel, is currently calculable at LO (one-loop), while NLO QCD effects being only available for the off-shell case with approximate or even complete $m_{\rm top}$ dependence \cite{Caola:2015psa,Grazzini:2018owa,Grober:2019kuf,Alioli:2016xab,Buonocore:2021fnj,Agarwal:2024pod}.
In \reffi{fig:best_results_gg} we consider for the $\Pg\Pg$ contribution the same differential observables as in \reffi{fig:best_results}.
The ATLAS polarised predictions have been obtained \cite{ATLAS:2023zrv} from an unpolarised \sher~NLO-merged event sample reweighted according to LO \moca~predictions for polarisation fractions differentially in the polar decay angles \cite{Denner:2021csi}.
The ATLAS nominal distributions largely deviate from fixed-order LO predictions independently of the employed on-shell approximation. This can be traced back to the reweighting techique. First, the NLO-merged results are expected to embed a different polarisation-state balance compared to LO, mostly due to higher jet multiplicities which sizeably change the fractions of LO-suppressed polarisation modes (LL, mixed). This results in a departure of ATLAS simulations from LO both in the distribution shapes and in the normalisation of polarised cross sections, which is the case for the decay angle and the rapidity separation shown in \reffi{fig:best_results_gg}.
Second, the reweighting of unpolarised events has been carried out according to the differential description in one single observable, namely $\cos\theta_{\Pe^+}^*$ (see \reffi{fig:best_costhetastar_gg}), while a multi-differential reweighting is needed to have a proper description of other observables. Matching BW (\madg) LO predictions to PS goes in the direction of ATLAS simulations at the level of distributions shapes in $\cos\theta_{\Pe^+}^*$, while the normalisation factor remains off as expected. On the contrary, the matching of LO predictions in the NWA (\sher) to PS effects gives distributions which follow  more closely the fixed-order ones. The discrepancy between the two LOPS results can be traced back to the details of the matching scheme and of the non-perturbative effects. These results have to be taken with a grain of salt, as there is no unique prescription to model loop-induced $\PZ\PZ$ production beyond LO, owing to the overlap with higher-order QCD corrections to the $q\bar{q}$ and $q\Pg$ channels \cite{Caola:2015psa,Grazzini:2018owa,Grober:2019kuf,Alioli:2016xab,Buonocore:2021fnj,Agarwal:2024pod}. This especially concerns the treatment of massive fermion loops which give a large contribution to gluon-fusion production of longitudinal weak bosons \cite{Denner:2020bcz,Poncelet:2021jmj,Denner:2021csi,Denner:2023ehn,Dao:2023kwc}.

At variance with the main $q\bar{q}$ production channel, the overall picture of the modelling of polarised-$\PZ$ bosons in $\Pg\Pg$ fusion is poor and relies on reweighting techniques which could badly bias the polarised-template fit of LHC data. This situation demands an urgent input from the theory community to improve the perturbative description in QCD of the $\Pg\Pg$ channel. 

\section{Conclusions}\label{sec:concl}
We have presented precise Standard-Model predictions
for the LHC production of pairs of polarised $\PZ$ bosons in the 
four-lepton decay channel.

A detailed technical comparison of various Monte Carlo predictions has been carried out at fixed order in the QCD and electroweak (EW) coupling, up to next-to-leading-order (NLO) QCD and EW corrections,
highlighting percent-level uncertainties coming from the polarised-boson modelling in various on-shell approximations (pole, narrow-width and decay-chain). Both non-resonant and interference contributions are found at the percent level.

Higher orders in QCD are assessed including parton-shower (PS) effects via matching and merging to exact or approximate NLO QCD predictions: moderate differences 
are found in transverse-momentum distributions, while fair agreement amongst various approaches
is found at the level of angular-observable shapes. Approximate NLO merging performs well in
reproducing the transverse-momentum distribution tails predicted at next-to-next-to-leading-order (NNLO)
in QCD, especially for longitudinal-boson signals.

The first combination of NNLO QCD and NLO EW corrections is carried out for doubly polarised signals, providing the new state-of-the-art predictions for integrated and differential cross sections, as well as for joint polarisation fractions. Predictions at leading order (LO)
for the $\Pg\Pg$ loop-induced channel are also included.

QCD-scale uncertainties are evaluated with seven-point variations of the renormalisation and factorisation scales. 
Our best predictions for polarised fiducial cross sections obtained with additive or multiplicative combination of
QCD and EW corrections feature 2\% (4\%) QCD-scale uncertainties when excluding (including) $\Pg\Pg$ loop-induced contributions.

PS-matched and fixed-order predictions have been compared to the ATLAS-simulation results used in
the most recent $\PZ\PZ$ polarisation analysis with Run-2 data. A fair agreement has been found in the dominant production mechanism
for most of the angular observables at the level of distribution shapes, while the normalisation factors differ by missing
PS-matching/merging effects and NNLO corrections (which were missing at the time of the analysis).
The reweighting techniques used by ATLAS to account for higher-order effects introduce more sizeable deviations 
from NLO-accurate predictions presented in this work for kinematic observables that probe suppressed phase-space regions (large rapidity separations, moderate-to-high transverse momenta). 

The theoretical modelling of the $\Pg\Pg$ loop-induced contributions to polarised $\PZ\PZ$, currently limited to LO matched to PS, 
requires urgent improvements both in the perturbative and in the non-perturbative description.\\

The numerical results generated for this work are available in {\sc Yoda} format \cite{Buckley:2023xqh} at the public webpage \url{https://github.com/multibosons/pol-ZZ-raw-data}.

\section*{Acknowledgements}
The authors acknowledge support from the COMETA COST Action CA22130 which promoted this work. They are especially grateful to Ilaria Brivio for the coordination of the COMETA network, and to Ramona Gr\"ober for fostering the scientific activities of the COMETA Working Group 1.
We would like to thank Lucia Di Ciaccio, Iro Koletsou, Joany Manjarres, Matteo Presilla, Emanuele Re and Aonan Wang 
for fruitful discussions.
RC and GO acknowledge support from the Italian Ministry of University and Research (MUR) under the PRIN 2022 funding scheme, contract Nr.~2022RBYK7T.
AD and CH acknowledge support from the German Federal Ministry for Education and Research (BMBF)
under contracts Nr.~05H21WWCAA and Nr.~05H24WWA, and from the German Research Foundation (DFG) under
reference Nr.~DE~623/8-1.
MJ and RCLSA acknowledge partial financial support from the US Department of Energy under grant Nr.~DE-SC0010004.
MH acknowledges financial support from the COMETA STSM grant
Nr.~29f14d41.
DNL is funded by Phenikaa University under grant Nr.~PU2023-1-A-18.
GP acknowledges support from COMETA by means of the dissemination grants Nr.~0a330a08 and Nr.~42c99e4,
and from MUR through the research grant Nr. 20229KEFAM (PRIN2022, Finanziato dall'Unione europea - Next Generation EU, Missione 4, Componente 1, CUP H53D23000980006). 
KP is funded by the UK Research and Innovation (UKRI) Science and Technology Facilities Council (STFC) under grant agreement ST/T004568/2.
MS is funded by the Royal Society through a University Research
Fellowship (URF{\textbackslash}R1{\textbackslash}180549, URF{\textbackslash}R{\textbackslash}231031) and an Enhancement Award (RGF{\textbackslash}
EA{\textbackslash}181033, CEC19{\textbackslash}100349, and RF{\textbackslash}ERE{\textbackslash}210397) as well as the STFC grants
(ST/X003167/1 and ST/X000745/1).
MJ and RCLSA  acknowledge computational support from the University of Massachusetts Amherst Research Computing at the Massachusetts Green High Performance Computing Center.
JL, GP and GZ acknowledge computational support from the Max Planck Computing and Data Facility (MPCDF).
Part of this work was performed using the Cambridge Service for Data Driven Discovery (CSD3), part of which is operated by the University of Cambridge Research Computing on behalf of the STFC DiRAC HPC Facility (www.dirac.ac.uk). The DiRAC component of CSD3 was supported by STFC grants ST/P002307/1, ST/R002452/1 and ST/R00689X/1.
Many authors 
acknowledge the hospitality of the University of Toulouse where relevant parts of this project were carried out.
MH thanks Durham University for hosting her during the finishing
stages for part of this work.

\bibliographystyle{JHEP}
\bibliography{polvv}

\end{document}